\documentclass[IOP]{emulateapj}
\usepackage{epsf}
\usepackage{amsmath}
\usepackage{multirow}
\usepackage{rotating}
\begin{document}
\slugcomment{Accepted to ApJ on 16 Nov 2012}
\shortauthors{A. L. King et al.}
%\shorttitle{}
\title{Regulation of Black Hole Winds and Jets across the Mass Scale}
\author{A.~L.~King\altaffilmark{1},
J.~M.~Miller\altaffilmark{1},
J.~Raymond\altaffilmark{2},
A.~C.~Fabian\altaffilmark{3},
C.~S.~Reynolds\altaffilmark{4},
K.~G\"ultekin\altaffilmark{1},
E.~M.~Cackett\altaffilmark{3,5},
S.~W.~Allen\altaffilmark{6,7},
D.~Proga\altaffilmark{8,9},
T.~R.~Kallman\altaffilmark{10}}

\altaffiltext{1}{Department of Astronomy, University of Michigan, 500
Church Street, Ann Arbor, MI 48109-1042, ashking@umich.edu}
\altaffiltext{2}{ Smithsonian Astrophysical Observatory, 60 Garden Street, Cambridge, MA 02138, USA}
\altaffiltext{3}{Institute of Astronomy, University of Cambridge, Madingley Road, Cambridge, CB3 OHA, UK}
\altaffiltext{4}{Department of Astronomy, University of Maryland, College Park, MD 20742, USA}
\altaffiltext{5}{Wayne State University, Department of Physics and Astronomy, Detroit, MI 481201, USA}
\altaffiltext{6}{Kavli Institute for Particle Astrophysiccs and Cosmology, Department of Physics, Stanford University, 452 Lomita Mall, Stanford, CA 94305, USA}
\altaffiltext{7}{SLAC National Accelerator LAboratory, 2575 Sand Hill Road, Menlo Park, CA 94025, USA}
\altaffiltext{8}{Department of Physics, University of Nevada, Las Vegas, NV 89154, USA}
\altaffiltext{9}{Princeton University Observatory, Peyton Hall, Princeton, NJ 08544, USA}
\altaffiltext{10}{Laboratory for High Energy Astrophysics, NASA Goddard Space Flight Center, Code 662, Greenbelt, MD 20771, USA}
\label{firstpage}

\begin{abstract}
We present a study of the mechanical power generated by both winds and jets across the black hole mass scale. We begin with the study of ionized X-ray winds and present a uniform analysis using {\it Chandra} grating spectra. The high quality grating spectra facilitate the characterization of the outflow velocity, ionization and column density of the absorbing gas. We find that the kinetic power of the winds, derived from these observed quantities, scales with increasing bolometric luminosity as $\log (L_{\rm wind,42}/C_v) = (1.58 \pm 0.07) \log (L_{\rm Bol,42}) -(3.19 \pm0.19)$. This suggests that supermassive black holes may be more efficient than stellar-mass black holes in launching winds, per unit filling factor, $C_v$. If the BHB and AGN samples are fit individually, the slopes flatten to $\alpha^{BHB}=0.91\pm0.31$ and $\alpha^{AGN} = 0.63\pm0.30$ (formally consistent within errors). The broad fit and individual fits both characterize the data fairly well, and the possibility of common slopes may point to common driving mechanisms across the mass scale. For comparison, we examine jet production, estimating jet power based on the energy required to inflate local bubbles. The jet relation is $\log (L_{\rm Jet,42})= (1.18 \pm 0.24) \log (L_{\rm Bondi,42}) - (0.96 \pm 0.43)$. The energetics of the bubble associated with Cygnus X-1 are particularly difficult to determine, and the bubble could be a background supernova remnant. If we exclude Cygnus X-1 from our fits, then the jets follow a relation consistent with the winds, but with a higher intercept, $\log(L_{\rm Jet,42}) = (1.34 \pm 0.50) \log (L_{\rm Bondi,42}) - (0.80\pm0.82)$. The formal consistency in the wind and jet scaling relations, when assuming ${\rm L_{Bol}}$ and ${\rm L_{Bondi}}$ are both proxies for mass accretion rate, suggests that a common launching mechanism may drive both flows; magnetic processes, such as magneto-hydrodynamics and magnetocentrifugal forces, are viable possibilities. We also examine winds that are moving at especially high velocities, $v >0.01c$. These ultra-fast outflows tend to resemble the jets more than the winds in terms of outflow power, indicating we may be observing a regime in which winds become jets. A transition at approximately $L_{\rm Bol} \approx 10^{-2} L_{\rm Edd}$ is apparent when outflow power is plotted versus Eddington fraction. At low Eddington fractions, the jet power is dominant, and at high Eddington fractions, the wind power is dominant. This study allows for the total power from black hole accretion, both mechanical and radiative, to be characterized in a simple manner and suggests possible connections between winds and jets. X-ray wind data and jet cavity data will enable stronger tests.
\end{abstract}

\section{Introduction}
Both winds and jets are thought to be driven by accretion disks; jets may be launched from the innermost regions, while winds may originate further out in the accretion disk \citep[e.g.,][]{Blandford82,Proga03}. It remains to be seen just how winds and jets relate. Are winds and jets driven by similar mechanisms? Does one quench the other? What role does the geometry of the magnetic field lines play?
One might expect the same physical launching mechanisms across the mass scale. Jet production is predominantly ascribed to magneto-hydrodynamics (MHD) across all mass scales; whether it is purely through MHD in the disk \citep[e.g.,][]{Lovelace76,Blandford82}, or through the disk and black hole \citep[e.g.,][]{Blandford77,Krolik10}. However, it has yet to be shown how wind properties scale with mass. Does radiation driving, thermal driving or magnetic processes drive these winds? 

``Warm-absorbing" winds detected as X-ray absorption features are seen in up to 50\% of Active Galactic Nuclei (AGN) \citep{Reynolds97,George98} and in the soft spectral state in stellar-mass black holes \citep{Miller06b,Miller06,Miller08,Ueda09, Neilsen09,King12a,Ponti12}. These blue-shifted absorption features are highly ionized and can potentially probe regions close to the black hole and basic disk physics. 

The winds observed in both black hole binaries (BHB) and supermassive black holes (SMBH) are capable of removing enormous amounts of material, even exceeding the mass accretion rates \citep[e.g.,][]{Blustin05,King12a}. They are typically wide angle outflows moving at a few hundreds of km s$^{-1}$. Jets, on the other hand, are highly collimated and have a much higher outflowing velocity, i.e. near the speed of light. Consequently, the mechanical energy in these jets can be much higher than in winds. In addition, using only the radiative luminosity in jets severely underestimates the power released in these systems because a majority of the energy is mechanical \citep[e.g.,][]{Gallo05,Churazov05,Allen06,Merloni07}. More importantly, strong mechanical feedback from a black hole can have a significant impact on its surroundings, including galactic formation, structure and co-evolution \citep[e.g.,][]{Kauffmann00,Churazov02,Croton06,Ostriker10,Gaspari11,Fabian12}.

On the low end of the black hole mass scale, the driving mechanisms of BHB X-ray winds are generally ascribed to either thermal pressure or magnetic mechanisms \citep{Begelman83,Woods96,Proga03}. Absorption features of highly ionized Fe XXV and Fe XXVI are the most commonly detected, implying the ionization parameters of the gas in BHB are very high, i.e. $\log \xi > 3$ \citep{Miller06,King12a,Ponti12}. At such high ionization states, line driving from the radiation field is inefficient at accelerating the winds to high velocities \citep[e.g.,][]{Proga02}.
On the high mass end, X-ray winds from SMBH systems span a larger range in ionization ($0 < \log \xi < 5$). Therefore, not only are Compton heating and magnetic mechanisms plausible driving mechanisms, but radiation pressure at low ionizations is also a plausible driving mechanism \citep{Proga00}. 

In this paper, we begin to examine the mechanical outflow power released in both winds and jets. In addition, we examine the outflow power across the mass scale, including stellar-mass and supermassive black holes. This will permit a complete characterization of the output of black holes: both radiative and mechanical. Characterization of the mechanical power can be particularly important on larger scales with respect to AGN feedback. AGN outflows, both jets and winds, may be responsible for shaping their environment, whether providing a source of hot ionized gas or by influencing the stellar velocity dispersion as evidenced by the $M-\sigma$ relation \cite[e.g.,][]{Gultekin09b}. These outflows may also play a vital role in the growth of black holes \cite[e.g.,][]{Hopkins05,Loeb05}. 

We describe how we derive the samples in Section \ref{sample}. Then, we estimate the kinetic power generated by winds and jets in these systems in Section \ref{methods}. Next, in Section \ref{results}, we describe the results for both the wind and jet relations, while in Section \ref{discussion} we present the conclusions as well as context for this study. We assumed $H_0$ = 70 km s$^{-1}$ Mpc$^{-1}$, $\Omega_\Lambda = 0.7$, $\Omega_M =0.3$ throughout this work. All errors are 1$\sigma$ statistical uncertainties unless otherwise stated. 

\section{The Mechanical Outflow Sample}
\label{sample}
\subsection{ X-ray Winds}
The central goal in assembling this wind sample is to create a uniform, unbiased and cohesive set of standards to ensure high quality spectra and rigorous results. These requirements are: (1) blue-shifted X-ray absorption features, (2) {\it Chandra} grating spectra, (3) photoionization modeling, (4) at least a 3$\sigma$ significant detection, and (5) velocity outflows of less than $~$3,000 km s$^{-1}$.

Wind and jet launching mechanisms are of prime interest in this analysis,
so it is important to probe the wind regions that are likely to be closest
to the base of the flow, and closest to the black hole. This points to a
comparison of X-ray winds across the mass scale. Pragmatic considerations
also make this the only consistent comparison that can be drawn. In AGN,
UV radiation pressure may help to accelerate winds (but radiation pressure
may not be sufficient to lift gas off of the disk in the first place);
however, UV winds are not observed in BHB's, likely owing to the high
ionization parameters found in these winds ($3 \leq \log \xi \leq 5$).
Fortuitously, restricting our analysis to X-ray winds in AGN captures the
bulk of the mass outflow rate. Prior treatments of ``warm absorbers" in
AGN have found that the mass outflow rate scales with the ionization
parameter estimated in different components \citep[e.g.,][]{Blustin05,Crenshaw12}.
We also require that observations of the X-ray winds be made with the
gratings spectrometers aboard {\it Chandra}. Although the XMM-Newton reflection
grating spectrometer (RGS) is similar in many respects, it covers a
narrower energy range, and it has a lower spectral resolution. The higher
resolution of the Chandra gratings means that it is more sensitive to lines
that are intrinsically narrow and weak, because more line flux is
concentrated in fewer spectral bins. Thus, this selection criterion serves
the aim of not biasing our result against weak or slow X-ray winds.
Similarly, we do not consider CCD spectra of X-ray winds in our initial
analysis since the modest resolution of such data inhibits the detection of
weak lines and modest velocity shifts.

In addition, we require that the winds be observed as blue-shifted absorption features with respect to the host. A significance of at least 3$\sigma$ for each component after fiducial fits with a canonical Galactic absorption and a power-law is also required. Further, to quantify these particular features, we demand that self-consistent photoionization modeling be performed to determine the gas parameters: the outflowing velocity, $v_{\rm out}$, and the ionization parameter, $\xi$. The ionization parameter is defined as $\xi = L_{ion} (nr^2)^{-1}$, where $L_{ion}$ is the ionizing luminosity, $n$ is the density, and $r$ is the distance from the warm absorber to the ionizing source. There are also references in the literature to the ionization parameter $U$, which is defined as $U=Q(4\pi nr^2 c)^{-1}$, where Q is the number of ionizing photons and $c$ is the speed of light. We use $nr^2=Q (4\pi U c)^{-1} = L_{ion} \xi^{-1}$ to convert to the ionization parameter $\xi$. It should be noted that $\xi$ depends on the X-ray continuum which is well constrained by observations, where as $U$ depends on the number of ionizing photons, which is very model dependent. The physical characterization of ionization in the spectra relies on multiple lines to determine the column density, ionization state and velocity shifts of the wind components. We include fits published in the literature. A majority of those fits were obtained with XSTAR grids \citep{Bautista01}, which mainly use $\xi$ to characterize the ionization of the gas. However, some fits were also obtained with CLOUDY grids \citep{Ferland98}, which mainly use $U$ to characterize the ionization of the gas. 

Finally, we restrict the outflowing velocity to be less than $~$3,000 km s$^{-1}$ (0.01$c$). We reserve outflows with a velocity faster than this for a sub-sample of tentative ultra-fast outflows. As these fast winds approach such velocities they may resemble jets more so than the typical lower velocity outflows. 

As a result of these criteria, we select thirteen AGN and ten BHB observations. Table~\ref{tab:indiv} lists the ionization, velocity and kinetic power for each observed outflowing component that is used in this work. Table~\ref{tab:lum} reports the total kinetic luminosity from each observation, summing over all the outflowing components. Figure~\ref{fig:vel} plots the observed velocity shift as compared to the ionization parameter, and Figure~\ref{fig:indiv} plots each individual absorption component's kinetic luminosity per filling factor, $C_v$, versus bolometric luminosity. 
We note here that our sample may suffer from a few selection biases. The first being that we may be biasing ourselves to high luminosity sources for which we get the best signal to noise. However, our sample does not necessarily include just the brightest sources, but also sources that are relatively faint and have long exposures in order to increase their signal to noise. (See the following sections for details of particular observations.) Therefore, a luminosity bias should not play a major role in our data. In addition, as will be shown in Section \ref{methods}, the kinetic wind luminosity does not directly depend on the column density, i.e. signal to noise, but instead on the velocity and ionization of the gas. Therefore, as long as a significant detection is made, the depth of the absorption features will not serve to bias our samples. Another bias could be the exclusion of {\it XMM-Newton} data because of its poorer resolution. However, as noted previously, the lower resolution of the RGS would serve to bias the sample against weak features. Further caveats are discussed in Section \ref{Caveats}.

\subsection{Supermassive Black Holes}
This sample, seen in Table \ref{tab:indiv}, is predominately derived from the work by \cite{Mckernan07}, which is comprised of 15 nearby AGN that were all observed with {\it Chandra} High Energy Transition Grating Spectrograph (HETGS) before July 1, 2003. The summed MEG first-order spectra were used. For further details of the data reduction, we refer the reader to \cite{Mckernan07} and \cite{Yaqoob03}. 

Owing to the observational selection criteria, only 7 of the 15 AGN show statistically significant blue-shifted absorption features. \cite{Mckernan07} perform a uniform analysis with XSTAR models, which characterizes the absorption and emission features seen in the sampled spectra. The XSTAR models were generated assuming an individual SED for each AGN. Each grid of models had an assumed density of $n_e = 10^{8}$ cm$^{-3}$ and a turbulent velocity of 170 km s$^{-1}$. These AGN span a redshift range of z=0.003 to z = 0.046, a mass range of 6$\times 10^5 - 10^8$ M$_\odot$, and a range of environments from centers of clusters to field galaxies. 

A majority of these AGN had more than one out-flowing component. See Table~\ref{tab:indiv}. These components were separated not only in velocity space but in ionization parameter as well. In these cases, we took the sum of all the components to evaluate the mass outflow rate, $\dot{M}_{wind}$, and kinetic wind luminosity, $L_{\rm wind}$, so as to compare the total out-flowing material and consequently the total power generated by these winds. See Section \ref{methods} for details. The values for the kinetic wind power and bolometric luminosities are given in Table~\ref{tab:indiv} for individual components and Table~\ref{tab:lum} for the summed components. We have assumed a covering fraction, $\Omega$, of $\Omega = 2\pi$, and given the results per filling factor, $C_v$.

In addition, six other AGN observations are included in our sample: NGC 4051, NGC 4593, Mkn 509, IRAS 18325$-$5926, NGC 4151 and Mrk 290. These are all relatively nearby AGN, z = 0.002 -- 0.034, with comparable masses spanning M = 2 -- 160 $\times 10^{6} M_\odot$. 
\cite{King12b} report observations of the nearby, Seyfert-1 AGN, NGC 4051 which show evidence of warm-absorbers. They coadded 12 {\it Chandra} HEG and MEG spectra from November 2008 for a total exposure time of 308 ksec. Although, \cite{Mckernan07} report detections of warm-absorbers in NGC 4051, the work by \cite{King12b} uses a different data set which was observed over 5 years later. \cite{King12b} use both XSTAR and Cloudy photoionization codes to create grids of models to fit to the data. By modeling the spectra with two separate photoionization codes, they were able to determine that three different ionization components were required by the data, independent of the model used. For this analysis, we use the three components that were determined using the XSTAR photoionization grids, which span a wide range in ionization and velocity, i.e. $\log \xi = 1, 3.4, \& \ 4.5$ and $v = 400, 630, \& \ 680$.

We also used the work by \cite{Steenbrugge03}, who observed NGC 4593, both with {\it Chandra} LETGS and XMM-Newton. We used only the 108 ksec LETGS spectrum that was analyzed with an XSTAR model. The XMM-Newton observation, taken 7 months later, could not constrain the absorption component ionization; however, it was consistent with the LETGS observation. The statistically significant absorption component in the LETGS observation had an ionization of $\log \xi = 2.61\pm 0.09$ with an outflowing velocity of 400$\pm 121$ km s$^{-1}$.

\cite{Ebrero11} also used the {\it Chandra} LETGS to observe the AGN Markarian 509. This 180 ksec observation was modeled with {\it xabs}, created using Cloudy. The fit to the spectra resulted in 3 components, one of which was significant at the 3$\sigma$ confidence level. This had an ionization of $\log \xi =2.26\pm$0.07 and a velocity of $v=196^{+87}_{-73}$ km s$^{-1}$, and is included in our sample. 

In addition, we used the work by \cite{Zhang11b}, who observed IRAS 18325$-$5926 . The authors coadd two exposures from March 2003 to get a total {\it Chandra} HETG spectrum of 108 ksec. Using a grid of XSTAR models, they find two outflowing absorption components with typical warm absorbing parameters of $\log \xi = 1.58\pm0.09$ and $\log \xi = 2.35 \pm 0.25$ and v$_{\rm out}$ = 340$\pm110$ km s$^{-1}$ and v$_{\rm out}$ = 460$\pm 220$ km s$^{-1}$, respectively. We note that \cite{Mocz11} also get results consistent with the higher velocity component.

\cite{Kraemer05} describe an analysis of NGC 4151, a nearby Seyfert-2 (it is more like a Seyfert-1 in X-rays), using two coadded {\it Chandra} HETG spectra from May 2002. The total exposure time was 250 ksec. The focus of their work was to describe the absorption features seen in both the X-ray and UV using Cloudy models. \cite{Kraemer05} do not quote the significance of the detection of these features; however, the individual lines are detected at several times their minus-side errors, and modeled together, should be quite significant. The X-ray absorption components taken from \cite{Kraemer05} that are included in our study have ionization parameters of $\log U \approx -0.27$ and $\log U \approx 1.05$ and outflowing velocities of $v_{\rm out} \approx 500$ km s$^{-1}$.

Finally, the last AGN in our sample, Markarian 290, is taken from \cite{Zhang11}. We note these authors use both {\it Chandra} and {\it XMM-Newton} grating spectra in their analysis. However, we only include the {\it Chandra} HETG spectra in our analysis. This particular spectrum had three coadded observations giving an exposure time of 166 ksec. The two XSTAR grid components that were statistically significant spanned a range of ionization states from $\log \xi = 1.62\pm0.15$ to $\log \xi = 2.45\pm0.04$ and a range of velocities from $v_{\rm out} = 450\pm30$ to $v_{\rm out}=540\pm150$ km s$^{-1}$, respectively. 

\subsection{Stellar-Mass Black Holes}
To the greatest extent possible, values in the literature were used to
estimate the kinetic power in winds observed in stellar-mass black
holes. As with supermassive black hole winds, only observations obtained with high resolution
gratings spectra and some level of photoionization modeling are
included. As a result of these considerations, only {\it
Chandra}/HETG spectra were selected. In total, ten observations
from four stellar-mass black holes and black hole candidates are
included in our initial analysis. It should be noted that many more
HETG observations are available -- GRS 1915$+$105 has been observed
regularly -- but we have only included prominent low- and high-flux
spectra from GRS 1915$+$105 in order to represent the properties of
the group rather than just one source. The winds in X-ray binaries have 
only been detected in the ``high/soft" X-ray spectral state \citep[e.g.,][]{Ponti12}. 
We do not include upper limits when the sources are in the ``low/hard" state. See 
Section 5.2 for more discussion of spectral state dependence. 

{\bf GRO J1655$-$40}: This is a famous and recurrent transient, and
its mass and distance are well-determined \citep[${\rm M} = 7.0\pm 0.2~ {\rm
M}_{\odot}$ and $d = 3.2\pm 0.2$~kpc][]{Hjellming95,Orosz97}. The presence of density-sensitive Fe XXII
absorption lines in a {\it Chandra} spectrum (ObsID 5461) of GRO J1655$-$40 allowed
for direct constraints on the density of the disk wind in this source
\citep[${\rm log}(n) = 14$;][]{Miller08}. Fits with an
independent photoionization code, grids of Cloudy models, and grids of
XSTAR models are reported in \cite{Miller08}. Based on that
work, we have used values of $L = 5.0\pm 0.1 \times 10^{37}~ {\rm
ergs}~ {\rm s}^{-1}$, an outflow velocity of $v = 500~ {\rm km}~ {\rm s}^{-1}$, ${\rm
log}(\xi) = 4.9\pm 0.1$, and $\Omega = 2.5$ in estimating the
kinetic power of the wind in GRO J1655$-$40. The wind observed in GRO
J1655$-$40 is particularly complex, and no single velocity
characterizes all of the lines observed; $v = 500~ {\rm km}~ {\rm
s}^{-1}$ is a representative value that is used throughout \cite{Miller08}
because it captures the outflow well. The resulting kinetic power
is broadly consistent with numbers given in \cite{Miller08}
estimated using the wind density. We note that \cite{Neilsen12}
also model a {\it Chandra} observation (ObsID 5460) that was made a month prior to this observation (ObsID 5461).
We find evidence for an absorption feature at 6.97 keV, which is consistent with no outflow, and therefore do not include it in this analysis.

{\bf H 1743$-$322}: This source is also a well-known and recurrent
transient \citep[see, e.g.,][]{Homan05,Corbel05, Miller06, Miller-Jones12}. The high column density
along the line of sight to H 1743$-$322 has permitted the detection of
a counterpart \citep{Steeghs03} but has not permitted a radial
velocity measurement. In this work, we have assumed a distance of
8.5$\pm 0.8$ kpc \citep{Steiner12}, and a
fiducial mass of 10~${\rm M}_{\odot}$. During its 2003 outburst, a
disk wind was clearly detected in thwo {\it Chandra}/HETG
observations \citep{Miller06}; parameters obtained from
photoionization modeling of each spectrum are used in this analysis.

In particular, the broadband X-ray spectral fits in table 2 of \cite{Miller06} were used
to derive bolometric luminosities, and the ionization parameters given
in table 5 in \cite{Miller06} are used. The first observation included has an
outflowing velocity of 670$\pm$170 km s$^{-1}$, while the second observation
included has an outflowing velocity of 340$\pm$170 km s$^{-1}$. The ionization
of the two observations are roughly the same at $\log \xi = 5.5\pm0.1$ and $\log
\xi = 5.6 \pm 0.1$, respectively. Less is known about the binary parameters of H
1743$-$322 than e.g. GRO J1655$-$40, and a larger value of $\Omega =
2\pi$ was adopted in calculating the photoionization models \citep[the
package used was an update of the code described in][]{Raymond93}.

{\bf GRS 1915$+$105}: This is an extremely well-known
microquasar. The mass of the black hole and its distance have been
determined \citep[${\rm M} = 14\pm 4~ {\rm M}_{\odot}$; $d = 12.5$~kpc;
][]{Greiner01}. A long {\it Chandra}/HETGS observation of GRS
1915$+$105 in a soft phase was analyzed in detail by \cite{Ueda09}; 
some simple photoionization modeling techniques
were applied to describe the disk wind that was detected. The
broadband spectral fits detailed in that work were used to calculate a
bolometric luminosity for this observation ($L = 7.5\pm 0.8 \times
10^{38}~ {\rm ergs}~ {\rm s}^{-1}$). As with the HETGS observation of
GRO J1655$-$40 reported in \cite{Miller08}, this spectrum of GRS
1915$+$105 is particularly rich, and no single velocity can describe
all of the ions observed. In this work, we adopt a value of $v = 300~
{\rm km}~ {\rm s}^{-1}$ because it is consistent with many ions and
achieves a balance between the most and least ionized components of
the flow. The line properties and analysis reported in \cite{Ueda09}
are consistent with $\Omega = 2.5$ and
${\rm log}(\xi) = 4.3$ (errors are not reported). We have used these
values in calculating the kinetic power of the wind in this
observation.

\cite{Neilsen09} treat a number of {\it Chandra}/HETGS observations of
GRS 1915+105. \cite{Ueda09} focused on the
observation called ``S1" in the \cite{Neilsen09} scheme; it is the
lowest luminosity of five ``soft" observations considered in their work. In
each soft observation, an ionized X-ray disk wind is detected at high
significance. To understand the range of wind properties in this important
source, then, we have included observations S2--S4 in this analysis.

Reduced and calibrated spectral and response files for each observation
were obtained through the {\it Chandra} ``tgcat" facility \citep[for details of
the observations, please consult ][]{Neilsen09}. The combined
first-order HEG spectrum from each observation was fit in the 2.3--9.0 keV
band with a simple phenomenological model consisting of disk blackbody and
power-law components. The lower energy bound was set by the high column
density along the line of sight to GRS 1915$+$105 ($N_{H}$ was fixed at
$5.0\times 10^{22}~ {\rm cm}^{2}$ in each case); the upper bound was set by
the likely calibration residuals in the high energy portion of the spectra.
This continuum model is not unique, but it allows for a simple and
accurate characterization of the flux in the observed energy band, and
gives reasonable basis for extrapolating to the 0.5-10.0 keV band.

We then calculated and applied grids of XSTAR models. A customized grid was
made for each observation, using the observed spectral continuum and the
unabsorbed 0.5--10.0~keV luminosity as the spectral input (the power-law
was truncated below 1~keV to prevent unphysical results). Again,
additional details of this procedure can be found in \cite{Miller06b,Miller08} 
and \cite{King12b,King12a}. In all cases, a turbulent velocity of
$500$~km/s, an iron abundance of 2.0 times the solar value \citep[e.g.][]{Ueda09},
a density of ${\rm log}(n) = 12.0$, and a covering
factor of $\Omega = 2.5$ were assumed.
In estimating the kinetic power in the wind for observations S2--S5, we
used the velocity shifts reported by \cite{Neilsen09}, and the
ionization parameters measured through direct fits to the {\it Chandra}
spectra with the XSTAR models. The values for all relevant parameters are given
in Table~\ref{tab:lum}.

{\bf 4U 1630$-$47}: Last, we considered an archival {\it Chandra} HETG
observation of the black hole candidate 4U 1630$-$47. Observation
4568 started on 2005 August 9 at 20:16:02 (UT), with a duration of
50 ksec. The instrumental configuration and modes used were the
same as those described in \cite{Miller08}. We again downloaded the
calibrated first-order gratings spectra and responses using the {\it
Chandra} ``tgcat" facility, and generated a combined first-order
HEG spectrum.

Fits to the continuum with an absorbed disk blackbody plus power-law
model give a high column density (N$_H$ = 7.8$\times 10^{21}$
cm$^{-2}$), a fairly hot disk ($kT = 1.36\pm 0.01$ keV), and a
power-law index of $\Gamma = 2.00\pm 0.02$. (The power-law index was
checked by making fits to a simultaneous {\it RXTE} PCA spectrum over
the 3--30 keV band.) Assuming a distance of 8.5 kpc, this continuum
model gives a bolometric luminosity of $2.2\pm0.2 \times10^{38}$
ergs s$^{-1}$. We further assume a mass of $10~ {\rm M}_{\odot}$ for 4U
1630$-$47 in this work.

Again as per the procedure in \cite{Miller08}, this spectrum was used
to illuminate gas in a grid of XSTAR models. Solar abundances were
assumed, and a turbulent velocity of 500 km s$^{-1}$ was found to give the
best fits to the data. As with H~1743$-$322, the parameters of the
binary system are not well known, and a covering factor of $\Omega =
2\pi$ was assumed in the generating the photoionization models. A
fiducial density of ${\rm log}(n) = 12$ was also assumed in generating
the models. As per the high luminosity observation of GRS 1915+105,
only an H-like Fe XXVI line was detected, immediately indicating a
high ionization. Direct fits with the XSTAR grid give an ionization
of ${\rm log}(\xi) = 4.9\pm0.4$, and a blue-shift of $v = 300\pm
200~ {\rm km}~ {\rm s}^{-1}$. 

\subsection{Jet Power}
\label{sec:jsample}
In collecting a jet sample, we also wanted to create a set of uniform standards and conditions that would ensure high quality and rigorous results just as we had done for the wind sample. Jets are found in the radio as a result of synchrotron radiation, so it is tempting to utilize the radiative portion of the energy as an estimate for the jet power \citep[e.g.,][]{Merloni07}. However, the majority of the energy carried off by the jets appears to be mechanical, not radiative \citep{Heinz02, Dimatteo03, Gallo05,Allen06, Taylor06}. We note that \cite{Merloni07} and \cite{Cavagnolo10} do find a relation between radio emission and mechanical power in jets, but a direct determination of the power is preferred over a proxy such as radio luminosity. In addition, \cite{Cavagnolo10} report the jet power relation to have a scatter of 0.70 dex and to be calibrated to high luminosity sources. It is unclear if extrapolation to lower luminosity sources is applicable. Likewise, the radio luminosities of jets are also subject to Doppler boosting \citep{Urry91}, which can be difficult to de-project, since the intrinsic spectrum must be known. We therefore restrict ourselves to only the most tentative of comparisons to radio luminosity via the fundamental plane of accretion onto black holes \citep{Merloni03,Falcke04,Gultekin09}. This plane relates the radio luminosity of SMBH to the accretion rate (via X-ray luminosity) and mass of the black hole. See Section \ref{indivmass} for further discussion. 

For a more stringent comparison with our wind sample, we require that the jet power be a direct estimate of the mechanical energy, not an indirect estimate using the radiative luminosity as a proxy for jet power. One way of quantifying the amount of mechanical power released via jets is to look at the volume they carve out in the form of ``cavities" or ``bubbles". These cavities are seen in both the radio and X-ray wavelengths. The energy ($E_{\rm Jet}$) is then estimated to be the sum of the internal energy and the $PdV$ work done to inflate the bubble. The time needed to carve out such a region is estimated at $t_{age}$=R/c$_s$, where $t_{age}$ is the age of the bubble, $R$ is the distance from the black hole to the center of the cavity and $c_s$ is the sound speed, typically estimated using X-ray observations. There are a number of different ways to estimate the age of the bubble, but using sound speed and bubble radius is a fine approximation as long as the bubble is still ``attached'' and is not buoyantly rising \citep{Dunn04}. Therefore, the power of the jet is $P_{\rm Jet} = E_{\rm Jet} / t _{age}$. We note here that this estimate is a long term average and is not sensitive to discrete episodes of jet emission, which would serve to increase the jet power estimate. 

For this work, we draw directly from the sample described in \cite{Allen06}, who use this prescription to analyze a sample of nine elliptical galaxies that display such X-ray and radio cavities. They also estimated the Bondi mass accretion rates by constructing radial temperature profiles close to the black hole from {\it Chandra} X-ray observations. As the \cite{Allen06} is a study of elliptical galaxies at low accretion rates, spherical accretion, i.e. Bondi accretion, is assumed with an efficiency conversion between mass accretion rate and luminosity \citep[$\eta = 0.1$ as given in][]{Allen06}. Conversely, in the wind sample, the accretion rates are typically higher at a few percent of Eddington, and the accretion is assumed to be through a standard thin disk. We use both the jet power and Bondi luminosities reported in \cite{Allen06}. We used the Bondi luminosities instead of the bolometric luminosities because each is an appropriate estimate for the mass accretion rate in these particular systems. \cite{Merloni07} report analysis of the same nine AGN as \cite{Allen06} as well as six additional sources. We do not include these extra sources in our sample as the temperature inside the Bondi radius was not measured directly but extrapolated from much further, outside regions. However, as will be shown in Section \ref{sec:jet}, \cite{Merloni07} report consistent analysis and results with \cite{Allen06}.

At the low mass end, it is much more difficult to use the same methods to estimate the jet power from bubbles and cavities. This is because most of the black hole candidates are not in regions with dense gas, making the bubbles hard to observe. The few that are embedded in dense clouds happen to also be in star forming regions. This means that the observed cavities can be carved out not only by their jets, but also by their high mass companion star's winds or even the supernovae associated with the black holes themselves. One such candidate is BHB, Cygnus X-1. This is a stellar mass black hole with an associated radio bubble that is thought to be created by its jet \citep{Gallo05,Russell07}. This is evidenced by the fact that the long axis of the bubble is aligned with the jet axis. However, there is no counter-jet seen, and the bubble may also be associated with a supernova remnant \citep{Russell07}. 

\cite{Gallo05} and \cite{Russell07} have both used the observed cavity to estimate the power of the jet, employing similar techniques as \cite{Allen06}. We note that in making density estimates of the emitting region, both \cite{Gallo05} and \cite{Russell07} assume the emission is Bremsstrahlung radiation. However, as \cite{Marti96} show, the radio emission in the limb brightened areas have steep spectra, which implies the regions are non-thermal in nature. Moreover, \cite{Russell07} find that the emitting loop is not
visible in the R band, indicating that the emission detected with an
H$\alpha$ filter, is in fact H$\alpha$ rather than Bremsstrahlung
emission, as assumed by Gallo et al. (2005).

One can use the observed H$\alpha$ flux to estimate the average density of
ionized gas (about 6 $\rm cm^{-3}$) and follow the method of \cite{Gallo05}
to determine the jet power, $P_{\rm Jet} =10^{34}-10^{38}$ ergs s$^{-1}$. 
As is obvious by the four orders of magnitude, there are large uncertainties that go into this calculation.
Clumpiness of the emitting gas would give an overestimate
of the average density, and if the ionized gas is indeed produced by a
100 km s${^{-1}}$ shock, it occupies a very thin sheet compared to the apparent
size of the emitting region. Second, the neutral fraction in the emitting region,
taken to be 98\% by \cite{Gallo05} and zero by \cite{Russell07},
is not well known. Finally, when shocked gas has cooled to the
point that H$\alpha$ emission is efficient, its pressure is probably
dominated by the magnetic field, so the sound speed should be replaced
by the fast mode speed, which is several times larger.

Instead, if we combine the intensity measurements of \cite{Russell07}
with the shock wave models of \cite{Raymond88} and the theory
of interstellar bubbles blown by a continuous energy input \citep{Castor75,Weaver77}, we can get a tighter constraint on the power estimate. 
\cite{Russell07} measured
an intensity in the [O III] $\lambda \lambda$5007, 4959 lines of approximately
$1.5\times 10^{-15}\rm ~ergs~cm^{-2}~s^{-1}$ per square arcsec after correction for
extinction, in a 2' section of a cut through the NE part of the shell.
They also measured [O III] to H$\alpha$ + [N II] ratios that indicate shock
speeds of 90-200 $\rm km~s^{-1} $ (their figure 7). Shock waves in that
range produce 0.87$\pm$0.2 photons in the [O III] lines per H atom that
passes through the shock \citep{Raymond88}. The shell is limb-brightened, and comparison
of the 2' thickness with the 5' radius indicates an enhancement factor of
2.4. Thus
\begin{equation}
1.7 \times 10^7 =2.4 * 0.87 * n_0 V_s /(4 \pi) \rm photons~cm^{-2}~s^{-1}~sr^{-1}
\end{equation}
\noindent
where $n_0$ is the pre-shock density and $V_s$ is the shock speed in units of cm$~s^{-1}$.
Thus $n_0 V_s$ = $1.0 \times 10^8$. The expression for the radius of a wind-blown
bubble in the intermediate stage \citep[when the shock is radiative as in the Cygnus X-1
bubble; equation 21 of][]{Weaver77} can be converted to
\begin{equation}
L_{37} = 7.7 \times 10^{-8} n_0 V_s V_{100}^2 R_5^2
\end{equation}
where $L_{37}$ is the jet luminosity in units of $10^{37}~\rm erg~s^{-1}$,
$n_0 V_s$ is in units of cm$~s^{-1}$, $V_{100}$ is the shock speed in units
of 100 $\rm km~s^{-1}$ and $R_5$ is the bubble radius in units of 5 pc. For a
shock speed of 90-200 $\rm km~s^{-1}$ and an average bubble radius of 4 pc, this
impies a jet luminosity of $4 - 20 \times 10^{37}$ $\rm erg~s^{-1}$.

There are two important caveats to keep in mind, both of which could lead to a severe
overestimate of the jet luminosity. First, O stars in the region, including the companion of Cyg X-1
itself, might contribute ionizing flux that we are assuming to come from the bubble shock.
Second, the shell could be a result of the explosion that created the Cyg X-1 black
hole, rather than the jet. The alignment of the shell with the jet direction suggests,
however, that the jet plays a significant role in the energetic of the bubble. In
addition, there is some uncertainty involved with the reddening correction. In view of the uncertainties in the relevant quantities, in the best method of estimating the power required to inflate the bubble, and the origin of the bubble itself, we use the full range of power estimates noted above. We note there are tighter constraints if the bubble is only associated with inflation by the jet and reddening is unimportant.

\subsection {Ultra-Fast Outflows}
\label{ufos}
Finally, we also defined a smaller subsample of winds that are moving faster than $~$3,000 km s$^{-1}$ (0.01c) relative to the systemic velocity. For this sample, we relax our standards for the AGN and include both {\it Suzaku} and {\it Chandra} imaging spectrometers. As we only discuss four examples of these outflows, (one BHB, one nearby quasar and two gravitationally lensed, higher redshift quasars), this additional data set is only meant to be illustrative not exhaustive. This sample still requires that the absorption features be at least 3$\sigma$ significance.

The first of these ultra fast outflows is the BHB J17091$-$3624. \cite{King12a} discus two {\it Chandra} HETG observations, one of which has clear absorption features above 6.9 keV. Using XSTAR photoionization grids, they model these features self-consistently and find an ionization of the absorbing gas to be $\log \xi = 3.3^{+0.2}_{-0.1}$, moving at $v_{\rm out} = 9600^{+400}_{-500} $ km s$^{-1}$. We also use the second component at a slightly higher ionization state $\log \xi = 3.9^{+0.5}_{-0.3}$ and velocity, $v_{\rm out} = 15,400 \pm400$ km s$^{-1}$. These are the fastest outflows observed from a BHB candidate, and they bear resemblance to some of the outflows seen in quasars \citep[e.g.,][]{Chartas02,Chartas07,Tombesi11}. 

\cite{Tombesi11} used the {\it Suzaku} X-ray Imaging Spectrometer (XIS) to observe 3C 111, an AGN at z = 0.0485. Of the three observations in their study, one showed evidence of an absorption line in the Fe K band. This observation had an exposure of 59 ksec. Using XSTAR grids with turbulent velocity of 3000 km s$^{-1}$, this one feature was fit with an ionization parameter of $\log \xi = 4.32 \pm 0.12$ and an outflowing velocity of $v_{\rm out} = 0.106 \pm 0.006 c$. 

\cite{Chartas02} observed the quasar APM 08279+5255 with the {\it Chandra} Advanced CCD Imaging Spectrometer (ACIS) and noted outflows in the absorption spectra. By using lensed quasars, one is able to probe outflows which would otherwise be too faint to observe. This quasar is at a redshift of $z = 3.91$. The spectra of APM 08279+5255 shows two features at 8.05 keV and 9.79 keV in the rest frame of the host galaxy \citep{Chartas02}. If these correspond to Fe XXV then the outflowing velocities would be $0.2c$ and $0.4c$, respectively. We utilized both components in this analysis.

Finally, \cite{Chartas07} described a gravitationally lensed quasar PG 1115+080, which is at a redshift of $z = 1.72$. The authors also used ACIS and notice absorption features in the host Fe K band. PG 1115+080 has prominent absorption features at rest frame energies of 7.27 keV and 9.79 keV; both of which are used in our analysis. Associating these features with Fe XXV gives velocities of $0.09c$ and $0.40c$. These features were not modeled with a photoionization model, but an assumed ionization of $\log \xi$ = 3.5 is taken as an estimate for the ionization parameter from XSTAR models \citep{Chartas07}. Bolometric luminosities are taken from \cite{Chartas07}, while estimates of the ionizing luminosities were taken as the rest frame X-ray luminosity from 0.2--10 keV from \cite{Dai04}. 

\section{Methods}
\label{methods}
After acquiring the sample, we calculate the mass outflow rate in the wind systems. This is done using simplified, order of magnitude estimates based on the expression for spherical wind. It is modified by both covering and filling factors to account for the fact that winds are not spherical outflows:
\begin{equation} 
\label{eq:mout}
\dot{M}_{\rm out} = \Omega \rho r^2 v C_v 
\end{equation}
where, $\Omega$ is the covering factor ( $0 < \Omega < 4\pi$), $\rho$ is the mass density ($\rho = \mu m_p n_e$), $\mu$ is the mean atomic weight assumed to be $\mu = 1.23 $, $m_p$ is the mass of a proton, $n_e$ is the electron density, $r$ is the radius from the ionizing source, $v$ is the out-flowing velocity, and $C_v$ is the line of sight global filling factor. As the winds may be clumpy and filamentary, they are likely to have a small filling factor. This expectation is based on the observed variability of the absorption lines \citep[e.g.,][]{Crenshaw03,Elvis04,Risaliti09} as well as density diagnostics \citep[e.g.,][]{King12b}. We note that variability can be due to both motion along our line-of-sight as well as the duty cycle of the wind. The short timescales of variability suggest that the variability is likely due to small filling factor and clouds moving across our line of sight rather then dissipation of the wind itself. Further, there is an inconsistency in the literature as to the actual filling factors, as there are few direct constraints on this quantity. This factor should vary between different sources as well as with ionization, $\xi$, but a range from 10$^{-5} $ to 1 is seen in the literature across the full mass scale \citep[e.g.,][]{Miller08,Mocz11,Zhang11,King12b,King12a}. Finally, as shown in \cite{Giustini12}, the exact nature of the wind may depend on the ionization, velocity and density of the wind, and may be quite uncertain. Therefore, we do not assume a value for the filling factor for any of the measurements but leave the kinetic energy luminosity in terms of the filling factor. 

Equation~\ref{eq:mout} can be rewritten in terms of observable quantities from the X-ray absorption features using $\xi = L_{ion} / n_e r^2$, where $\xi$ is the ionization parameter and $L_{ion}$ is the ionizing luminosity between 1 -- 1000 Ryd (1 Ryd = 13.6 eV). 
\begin{equation}
\dot{M}_{\rm out}=\frac{\mu m_p \Omega L_{ion} v C_v }{\xi}
\end{equation}
For consistency, we use $\Omega = 2 \pi $ for all the AGN sources, based on findings by \cite{Reynolds97} who found half of all Seyferts show evidence for warm absorbers. We have not assumed a filling factor, $C_v$, but have reported our results of kinetic luminosity per filling factor.
To convert the mass outflow rate to the kinetic energy carried away by the warm absorbing winds, i.e., power or kinetic luminosity, we use the following relation,
\begin{equation}
L_{\rm wind} = \frac{1}{2} \dot{M}_{\rm out} v^2 = \frac{\mu m_p \Omega L_{ion} v^3 C_v}{2 \xi}. 
\end{equation}
The total kinetic luminosity is the amount of mechanical energy that is carried away by the wind. It is important to understand this in the context of the total escaping energy of the black hole and accretion disk. This can be done via comparison to the radiative power released, i.e., the bolometric luminosity. Further, the bolometric luminosity is often considered a proxy for the mass accretion rate by assuming an efficiency conversion, $L_{\rm Bol} = \eta \dot{M} c^2$. The efficiency, $\eta$, is usually assumed to be 10\% \citep[e.g.,][]{Allen06,Vasudevan09,Fabian09}, which is consistent with the Soltan's argument \citep{Soltan82}. However, in reality it is likely to vary between sources and with Eddington fraction. In our study, the bolometric luminosities for the AGN are taken from broad-band spectral energy distribution (SED) fitting performed by \cite{Vasudevan09}. In the few instances that the AGN lack a bolometric luminosity, we used the conversion $L_{\rm Bol} \approx 20 L_{2-10keV}$ \citep{Vasudevan09}. For the stellar-mass black holes, their SED peaks in the X-ray. Therefore, the bolometric luminosities are taken from the X-ray observations as the luminosity between 0.5-10.0 keV. Where values in the literature were quoted for different energy bands, the luminosity was converted to the 0.5-10.0 keV band for consistency by extrapolating the given models within Xspec. We also note that any uncertainty in distance, which could effect the bolometric and ionizing luminosities as well as estimates of the outflowing velocity, are small as compared to the uncertainties in these measured quantities.

The jet power is calculated using the energy estimates of radio and X-ray cavities and age of the bubble as described in Section \ref{sec:jsample}, $P_{\rm Jet} = E_{\rm Jet} / t _{age}$. We again note that this estimate is a long term average and short, discrete episodes of jet emission would increase the jet power estimates.

\section{Analysis and Results}
\label{results}
\subsection{Bolometric Luminosity versus Wind Power}
\label{bol}
After acquiring a sample of BHB and AGN with estimates for the wind power, we begin to analyze how $L_{\rm Bol}$ relates to the total kinetic outflowing power in each system. In this initial analysis, we only include the lower velocity winds (not the jets, which we consider in Section \ref{sec:jet}, or the ultra-fast outflows, which we consider in Section \ref{sec:ufo}). Figure~\ref{fig:vel} shows the distribution of velocities and ionization parameters that are included in the kinetic power of the winds. Figure~\ref{fig:indiv} shows the kinetic wind luminosity as compared to the source bolometric luminosity for individual components in each observation, while Figure~\ref{fig:wind} shows the same plot but for the total kinetic wind luminosity for each observation. The stellar-mass black holes cluster at the lower luminosities, while the SMBH are found at the higher luminosities as expected. A positive correlation is apparent in the data set.

In Figure~\ref{fig:indiv} (as well as the following figures), we plot the kinetic luminosity per filling factor, $C_v$, versus the bolometric luminosity. 
Further, Figure \ref{fig:wind} shows the total kinetic power for each observation, which uses the sum of the individual components plotted in Figure \ref{fig:indiv}. We begin our analysis with the total kinetic power for each observation. Initially, we assume that there is a common relation between both the AGN and BHB. In Section \ref{indivmass}, we relax this assumption and characterize the two groups separately. 

To first characterize the trend given in Figure~\ref{fig:wind}, we utilize two correlation tests: a Spearman's rank test and a Kendall's $\tau_K$ test. We find a Spearman's rank coefficient $\rho_S$ = 0.89, with a null hypothesis probability (i.e., no correlation) at $p= 1.8 \times10^{-8}$, indicating a strong and positive correlation. The value for Kendall's coefficient is $\tau_K$ = 0.72 with $p=1.4\times10^{-6}$, also indicating a strong, positive correlation. 

Following this, we assume that the data can be described by the linear relation,
\begin{equation}
y = \alpha x + \beta
\end{equation}
where, $y = \log (L_{\rm wind,42}/C_v)$, $x =\log (L_{\rm Bol,42})$, and the subscript ``42" denotes the units $10^{42} \ {\rm erg \ s^{-1}}$. We then minimize the function,
\begin{equation}
\chi^2 \equiv \sum^N_{i=1} \frac{(y_i - \beta- \alpha x_i)^2}{\alpha^2\sigma^2_{xi}+\sigma_{yi}^2 + \sigma^2_0}
\end{equation}
to estimate $\alpha$ and $\beta$ \citep[e.g.,][]{Press92,Tremaine02}. Here $\sigma_{yi}$ and $\sigma_{xi}$ are the errors associated with the kinetic wind luminosity and the bolometric luminosity, respectively. The quantity $\sigma_0$ is the intrinsic scatter in the relation and is determined by ensuring the reduced $\chi^2$ is close to unity. We obtain $\alpha = 1.58 \pm 0.07$, $\beta = -3.19 \pm 0.19$ and $\sigma_0 = 0.68$, such that, 
\begin{align}
\log \left( L_{\rm wind,42}/C_v \right) = & \nonumber\\
(1.58 \pm 0.07) \log & \left(L_{\rm Bol,42} \right) -(3.19\pm0.19).
\end{align}

These parameters are listed in Table \ref{tab:para}. The large $\sigma_0$ implies that the intrinsic scatter in these measurements is dominant over the measurement errors. We can expect a high intrinsic scatter due to the high variability of each of these sources, especially because the observations used to derive the AGN bolometric and wind luminosities of individual sources were not made simultaneously. Environment may also play a large role in driving this scatter, evidenced by the fact that the larger scatter is associated with the SMBH measurements, which are located in dense groups and clusters to open field environments. This scatter may also be attributed to the bolometric correction applied to the X-ray luminosities of the AGN and not to the BHB which peak in luminosity in the X-ray band.

Finally, the scatter may be due to the inclusion of a range of ionization parameters, especially in the AGN. There appears to be a stratification in the power which may depend on the ionization parameters. This is shown in Figure \ref{fig:indiv} as the low ionization components tend to have higher powers as compared to the higher ionization parameters. Therefore, we also examine only the high ionization, i.e. $\log \xi >2$, components with an aim to better compare the same sample in both AGN and BHB. We fit the individual components with high ionization and find the slope flattens to $\alpha^{\log \xi>2} = 1.42 \pm 0.06$, $\beta^{\log \xi>2} = -(3.73\pm0.14)$ and a scatter of $\sigma_0^{\log \xi>2} =0.57$. These results are given in Table \ref{tab:para} and shown in Figure \ref{fig:indiv}.

\subsection{ Jet Power}
\label{sec:jet}
We next include the relation between the Bondi luminosity and jet power as estimated via the radio bubbles seen in elliptical galaxies and Cygnus X-1 (See Figure \ref{fig:jet} and Table \ref{tab:lum}). Again, the Bondi luminosity is used for the ellipticals instead of bolometric luminosity as in the wind sample, but can be thought of as a proxy for mass accretion rate just as the bolometric luminosity is at high accretion rates. For the jet sample, a high degree of correlation between the Bondi luminosity and the jet power is indicated, as initially noted by \cite{Allen06}. We find a Spearman's rank coefficient of $\rho_S = 0.95$ with probability $p=2.3\times10^{-5}$, and a Kendall's $\tau_K$ coefficient of $\tau_K =0.87$ with probability $p=4.9\times10^{-4}$. Clearly, there is a positive correlation for this data set as well.

As a correlation is quite apparent in this jet sample, we fit the data using the same technique as was used for Sections \ref{bol}. We find a $\alpha^{jet} = 1.18 \pm 0.24$, $\beta^{jet} = -(0.96 \pm 0.43)$, and a intrinsic scatter consistent with zero ($\sigma^{jet}_0 =0$). These parameters are listed in Table \ref{tab:para} for comparison with the wind parameters. This relation is also shown in red in Figure~\ref{fig:jet}. The reduced $\chi^2$ is quite small at $\chi^2/\nu = 0.11$. This is a result of the large uncertainty estimates on the Bondi luminosities. The error in Bondi luminosity is estimated using the uncertainty given in \cite{Allen06} as well as the uncertainty derived from the scatter in the $M-\sigma$ relation given by \cite{Gultekin09b}. These two uncertainties are added in quadrature, resulting in an uncertainty of approximately 0.62 dex.

As \cite{Allen06} and \cite{Merloni07} perform similar fits to exclusively the elliptical galaxies, we next exclude Cygnus X-1 from the fit and find $\alpha^{jet} = 1.34 \pm 0.50$, $\beta^{jet} = -(0.8 0\pm 0.82)$ and an intrinsic scatter also consistent with zero. See the orange line in Figure~\ref{fig:jet}. Our analysis is able to reproduce the results by \cite{Allen06}, who first published this sample. \cite{Allen06} find $B=0.77\pm0.20$, which is equivalent to $\alpha^{jet} = 1.30 \pm 0.34$ in our nomenclature. Our results are also consistent with \cite{Merloni07} who find a slope of $\alpha^{jet}=1.6^{+0.4}_{-0.3}$ when correlating $\L_{Kin}/L_{\rm Edd}$ to $L_{Bondi}/L_{\rm Edd}$. 

We note here that Cygnus X-1 nominally lies one or two orders of magnitude above the elliptical jet relation but is consistent within its large uncertainty. This begs the question whether the radio bubble seen around Cygnus X-1 is truly related to the black hole jet or in fact a chance alignment. Cygnus X-1 is located in a fairly active star forming region where massive young stars may be responsible for such a structure \citep{Reid11}. In fact, there have been X-ray winds associated with the Cygnus X-1 system, whether from the accretion disk or the companion O star is unclear. These may also have an effect on inflating the bubble, which would bring the power estimate down. In addition, as previously discussed, we do not see a counter bubble from the presumed counter-jet.

If we include Cygnus X-1 in the jet relation, then the slope of the jet relation is inconsistent with the wind relation by only 1.6$\sigma$. However, when we exclude Cygnus X-1 \citep[as it is plausibly associated with a supernova remnant ][]{Russell07} the slopes of the jet and wind relations are consistent within errors. Although the normalizations are different, a common slope might imply a shared driving mechanism. In Figure~\ref{fig:jet}, it is apparent that the jet and wind power normalizations are within a few orders of magnitude, especially at high luminosity. However, correcting for the filamentary and geometric structure of the winds via the filling factor will decrease the wind power normalization by 3 to 4 orders of magnitudes, demonstrating the greater efficiency of the jet power.

\subsection{Spectral State Dependence}
It is also interesting to compare how jet and wind power scale in terms of Eddington fraction to examine the accretion rate dependence. Figure~\ref{fig:Edd} shows both the wind and jet power per Eddington luminosity as compared to their Eddington fraction (i.e., bolometric luminosity or Bondi luminosity per Eddington luminosity). The AGN jet power is denoted in red, Cygnus X-1 jet power is denoted in orange, the AGN winds are denoted in black, and the BHB winds are denoted in blue. The solid lines are taken from \cite{Churazov05} and describe the outflow mechanical power (thick line) and radiative power (thin line) for AGN. They postulate that AGN should follow a similar evolution to their stellar-mass counterparts, in that they should have a strong jet dominated phase at low accretion rates, and little-to-no jet production at high accretion rates. This allowed them to present a model for AGN feedback and co-evolution with the host galaxies as a function of mass accretion rate. We assume an efficiency of $\eta = 0.1$ to compare our bolometric luminosity to their mass accretion rate.

A division in the outflow power is seen at approximately 10$^{-2} L_{\rm Edd}$ in our data set. See Figure \ref{fig:Edd}. Below this Eddington fraction, jets dominate and increase in power with bolometric luminosity. At higher Eddington fractions, the wind power dominates but decreases with Eddington fraction. We note that there is an observed dichotomy between the type of outflows seen in stellar mass black holes and their X-ray spectral state already described in the literature. In particular, winds are found in the ``high/soft" state, i.e., high mass accretion rate and Eddington fractions ($\gtrsim 10^{-2} L_{\rm Edd}$), and radio jets are observed in the ``low/hard" state, i.e. low mass accretion rates and Eddington fractions \citep{Miller06b,Miller08,Neilsen09}. Although our sample is not exhaustive, we do see illustration of a similar trend in the AGN sample where jets persist at low Eddington fractions and winds persist at high Eddington fractions. We note that in one AGN source, NGC 4051, there is evidence for simultaneous winds and jets \citep{King11}. However, the winds may very well dominate at this high Eddington fraction in NGC 4051. Regardless of the particular outflow seen, Figure \ref{fig:Edd} allows for the prediction of the outflow power as a function of Eddington fraction. {\it This will be vital to simulations of AGN feedback and co-evolution; both for matching predictions to observations as well as implementing sub-grid physics in cosmological simulations.}

\subsection{Ultra-Fast Outflows}
\label{sec:ufo}
There are a few wind sources that lie well above the wind relation in Figure~\ref{fig:wind} and much closer to the jet relation in Figure~\ref{fig:jet}. This is primarily because of their high velocities, i.e., $v>0.01c$, as $L_{\rm wind} \propto v^3 / \xi$, which increases their power to lie near the jet relation. In Figure~\ref{fig:ufo}, we include four ultra-fast outflows mentioned in Section \ref{ufos}. These are denoted in the black squares. The upper squares assume a global filling factor of unity. For comparison, we include estimates of the wind luminosity if the filling factor were as low as $C_v = 10^{-4}$ connected by a dashed line to the original higher estimate. Such a small filling factor is reasonable as it is consistent with their potential transient nature \citep{Tombesi11} as well as density diagnostics \citep{King12b}. We note that variability of these sources can be attributed to both movement across our line-of-sight as well as duty cycle. However, the timescales for variability are short compared to dissipative timescales, and are generally ascribed to filling factor and not duty cycle \citep[e.g.,][]{Elvis04,Crenshaw12,Risaliti09}. Regardless of the filling factor, these high velocity outflows tend to be much more efficient at their given bolometric luminosity as compared to the other winds. Therefore the ultra-fast outflows may resemble the jet relation, which is plotted in Figure~\ref{fig:ufo}. As the ultra-fast outflow power approaches that of jets, it suggests that we are seeing the transition from winds as they are being accelerated into jets. 

\subsection{Distance and Mass Dependence Diagnostics}
\label{indivmass}
When examining broad relationships, it is important to be wary of a rising trend with a slope of unity; this can indicate that a relation that is driven by a mutual dependence of a third, shared parameter. For example, in the wind relation, both the bolometric and ionizing luminosity are both proportional to the square of the distance, which may have an influence in driving the observed trend. However, we find evidence to the contrary. Not only is the observed slope in the wind relation is greater than unity, but a partial correlation test, described by \cite{Akritas96}, gives a low probability of $p=0.035$ that the wind relation is driven by a mutual dependence on distance. This estimate is derived from Kendall's partial $\tau_{K,p}$, which gives the Kendall's $\tau$ holding the third parameter, distance, constant. The values we find for the wind relation are $\tau_{K,p}= 0.270$, with an estimated variance, $\sigma_K$ = 0.128. Further, the rising trend in the winds is therefore dominated by the velocity and ionization as $L_{\rm wind} \propto v^3/\xi$, both of which are directly and independently constrained by observations. We also note that the jet sample is not driven by distance. This sample has an even smaller probability of $p<10^{-6}$ that distance is needed as a third parameter. This is derived from a Kendall's partial correlation test where $\tau_{K,p}$=0.86 and $\sigma_K$=0.13.

Similarly, we tested whether the relations are driven by the mass of the black hole. A partial correlation test of the wind sample using mass as a third variable gives a probability of $p=0.034$ that the relation is driven by a mutual dependence on mass. This probability was derived from a Kendall's partial correlation test where we found $\tau_{K,p}$ is 0.436 with $\sigma_K$ = 0.206. The jet sample is even less dependent on mass with a probability of $p=1.8\times10^{-3}$ ($\tau_{K,p} = 0.73$ and $\sigma_K = 0.24$), likely the result of the sample including only 1 BHB and massive ellipticals ($M\sim10^9 M_{BH}$). 

Although the probability in the wind sample is small, it does not rule mass out as a third parameter at a 3$\sigma$ level like the jet sample does. The nature of these winds across such a large mass scale has not been studied before. Consequently, we explore the potential for mass dependence in this data set.

We separately fit a linear relation to both the BHB and AGN data sets. We used the same linear regression as in Sections \ref{bol} $\&$ \ref{sec:jet}, minimizing the $\chi^2$ for the best fit parameters. We find the BHB sample had the best fit parameters of $\alpha^{\rm BHB} = 0.91 \pm 0.31$, $\beta^{\rm BHB} = -(5.58\pm 1.68)$ and an intrinsic scatter consistent with zero ($\sigma_0^{\rm BHB}$=0). The AGN sample had the best fit parameters of $\alpha^{\rm AGN} = 0.63 \pm 0.30$ and $\beta^{\rm AGN} = -(1.24\pm0.63)$ with an intrinsic scatter of $\sigma^{\rm AGN}_0 = 0.58$. Figure \ref{fig:indivmass} depicts the best fit relation for the BHB and AGN, and Table \ref{tab:para} lists these parameters for comparison with previous results. 

The slopes of the two individual fits are consistent with each other, although inconsistent with the initial fit to the wind sample at the 2.4$\sigma$ and 3.1$\sigma$ level for the BHB and AGN samples, respectively. However, in order to evaluate whether these parameters are truly inconsistent with the initial fit, we used a bootstrap method to resample the data and estimate the number of trials we would expect with slope $\alpha \geq 1.58$. Using N=10$^4$ trials, we found that in the BHB sample, 3.1\% of the trials gave a slope $\alpha \geq 1.58$. Similarly, in the AGN sample, $1.7\%$ of the trials gave a slope $\alpha \geq 1.58$. Although these parameters are formally inconsistent with the initial fit using the 1 $\sigma$ error bars, we can not rule them out at more than a 98.3\% confidence level. In addition, by fitting the data separately, we are introducing an additional three free parameters and thereby doubling the parameter space. Because we can not directly compare the $\Delta\chi^2$, as the $\chi^2/\nu$ is fit to be unity, we used a Bayesian analysis to determine which model better describes the data. 

The advantage to using Bayesian statistics is that it allows us to compare two different models of the same data without a reduced $\chi^2$ and without the same number of degrees of freedom. This is done via a Bayesian odds ratio, which compares the likelihood of each model over the entire parameter space. In our analysis, we assume a uniform prior distribution in slope ($\alpha \in [-10,10]$), intercept ($\beta \in [-12,2]$), and scatter ($\sigma_0 \in [0,3]$). We find that when comparing the likelihood of the single fit, ${\cal L}_1$, to the two individual linear fits, ${\cal L}_2$, the odds ratio was ${\cal O}_{1,2} = \frac{{\cal L}_1}{{\cal L}_2} = 0.17$. This means that the two linear fits are slightly favored over the single linear fit, suggesting that mass may have a role in this relation. In Section \ref{Caveats} we suggest further observations that would also help to distinguish between the two models. 

We next fit a plane to the data, which included mass as an additional parameter. A similar linear regression to that used in Section \ref{bol}, is used to fit a plane to the entire wind data set. The plane is described by,
\begin{equation}
Z = \alpha_p X + \beta_p Y +\gamma_p
\end{equation}
where $X=\log (L_{\rm Bol})$, $Y= \log(M_{\rm BH})$, $Z = \log (L_{\rm wind}/C_v)$, and $\gamma_p$ is the intercept. In order to find the best fit parameters, we minimized the function,
\begin{equation}
\chi^2 \equiv \sum^N_{i=1} \frac{(Z_i - \gamma_p - \beta_p Y_i- \alpha_p X_{i})^2}{\alpha_p^2 \sigma^2_{X,i}+\beta_p^2 \sigma_{Y,i}^2 + \sigma^2_{p,0}}
\end{equation}
where $\sigma_{X,i}$ is the $\log(L_{\rm Bol})$ scatter, $\sigma_{Y,i}$ is the $\log(M_{\rm BH})$ scatter, and $\sigma_{p,0}$ is the intrinsic scatter of the plane in the $Z$ direction. This is further discussed in \cite{Merloni03} and \cite{Gultekin09}

We found the data set is best fit by the parameters are $\alpha_p = 0.2 \pm 0.4$, $\beta_p =1.2\pm 0.3$ and $\gamma_p = 24.5 \pm 0.2$ with an intrinsic scatter of $\sigma_{p,0}=0.68$. Figure \ref{fig:fp} shows the best fit plane. 

Ideally, we would like to compare this plane to a plane that describes the jet cavities while including mass. However, the jet cavities are dominated by ellipitical galaxies with masses of approximately $M_{BH} \approx 10^{9} M_\odot$, and as demonstrated by the partial correlation test, the data show a relation that is independent of mass at the 99.82$\%$ confidence level. Although mass may be an important variable, the data may not span a wide enough parameter space to deduce its effects. However, previous studies that have used radio luminosity instead of X-ray cavities to study jet characteristics, have shown a mass dependence \citep[e.g.,][]{Merloni03,Falcke04,Gultekin09}. As we noted in Section \ref{sec:jsample}, using radio luminosity to study jet properties involves some uncertainties when converting between radio flux density to jet power \citep[e.g.][]{Merloni07}, and should be treated with caution. 

Therefore, we proceed with only a tentative comparison of our relation to the fundamental plane of accretion onto black holes. The wind plane parameters, $\alpha$ and $\beta$, are formally consistent with the fundamental plane parameters given by \cite{Gultekin09}, although the overall normalizations differ. See Figure \ref{fig:Gultekin}. The plane given by \cite{Gultekin09} relates the accretion rate of low-luminosity black holes via the X-ray luminosity and the mass of the central black hole to the radio luminosity of the compact radio source in the host galaxy. We note that we do not include a conversion between radio luminosity and jet power in this comparison, which may be necessary for comparison with our wind sample. In addition, the work by \cite{Gultekin09} uses X-ray luminosity and does not include the bolometric correction factors that we have used in our work. Both of these caveats should be examined in the future to better understand the connection and consistency between these two relations. Finally, we note that our wind coefficient describing the bolometric luminosity, $\alpha_p$, is consistent with the coefficient given by \cite{Merloni03}, and our mass coefficient, $\beta_p$, is only inconsistent with the \cite{Merloni03} coefficient at the 1.3 $\sigma$ level. Again, \cite{Merloni03} describe a similar plane between the mass, X-ray luminosity and compact radio emission in both stellar- and supermassive black holes.

\section{Discussion}
\label{discussion}
We have compiled samples of both X-ray winds and relativistic radio jets that span eight orders of magnitude in black hole mass. Each sample has uniform, rigorous selection criteria to ensure consistent comparisons between the various sources. In particular, we demand that the winds be detected through significantly blue-shifted absorption features seen in the X-ray band, and the jets be seen as X-ray bubbles or cavities. By including only the X-ray winds and jets, we aim to probe the outflows associated with the inner accretion disk. These flows may be driven by the accretion disk, and jetted outflows may also tap the spin of the black holes.

For comparison, we also examine the bolometric or Bondi luminosity of each source. In doing so, we find a relation that describes the entire black hole X-ray winds sample as $\log (L_{\rm wind,42}/C_v) = (1.58 \pm 0.07) \log (L_{\rm Bol,42}) -(3.19\pm0.19)$ and jet sample as $\log (L_{\rm Jet,42})= (1.18 \pm 0.24) \log (L_{\rm Bondi,42}) - (0.96 \pm 0.43)$. If we exclude Cygnus X-1 the relation becomes $\log(L_{\rm Jet,42}) = (1.34 \pm 0.50) \log (L_{\rm Bondi,42}) - (0.80\pm0.82)$. These relations suggest a common regulation scheme for winds and jets across the mass scale. We also find that when fit individually, the BHB and AGN wind samples have shallower slopes of $\alpha^{BHB} = 0.91\pm0.31$ and $\alpha^{AGN} = 0.63\pm0.30$, which are also consistent with each other within errors. Although, the two wind fits are preferred slightly over the entire sample fit, a ${\it common}$ slope between the BHB and AGN is required by the data regardless of the procedure used. 

\subsection{Plausible Outflow Driving Mechanisms}
Examining winds specifically, thermal, radiative and magnetic mechanisms are viable methods of driving winds. However, it is not clear that these mechanisms would {\it collectively} drive this relation in the same way as required by the data. We now turn to whether radiative, thermal or magnetic processes can drive the observed X-ray wind correlation individually. 

First we examine whether radiation pressure, and more specifically UV line driving, has enough force to launch winds. If we assume this occurs when the force from the lines, $F_{lines}$, exceeds that of gravity, $F_{grav}$, i.e. $F_{lines} > F_{grav}$, it yields the following UV luminosity, $L_{UV}$, criterion that $L_{UV} \mathcal{M}(t) > L_{\rm Edd}$ \citep[See equation 8 in][]{Proga02}. Here $\mathcal{M}(t)$ is the force multiplier \citep{Castor75}, and $L_{\rm Edd}$ is the Eddington luminosity. The force multiplier allows us to quantify the contribution of line driving in addition to electron scattering and is a function of the optical depth $t = \sigma_T \rho v_{th} \left| \frac{dv}{dr} \right|^{-1}$, $\rho$ is the density, $v_{th}$ is the thermal velocity and $\left| \frac{dv}{dr} \right|$ is the velocity gradient along the flow \citep{Castor75}. For most BHB, the strong X-ray radiation can highly ionize the gas, driving the $\mathcal{M}(t)$ to 1 at $\log \xi \approx 2$, and $\mathcal{M}(t)$ to 0.1 at $\log \xi \approx 3$ \citep{Proga00}. In addition, the BHB spectrum does not have a large relative contribution from the UV, due to the high disk temperatures. This also hinders line-driving of winds consistent with \cite{Proga02,Proga02c}. On the other hand, AGN spectra peak in the UV, and the AGN winds span a wider range of ionization parameters, suggesting that at low ionization parameters, $\log \xi < 2$, line-driving may be important, which is also consistent with the work by \cite{Proga02c}. This may partially account for the index of 1.58$\pm$0.07 in the initial fit that makes AGN more efficient wind producers. However, it still remains to be seen what drives the higher ionization states found in a majority of the AGN listed here. 

Thermal pressure is another plausible driving mechanism. Winds can be driven by thermal pressure if the temperature of the gas is higher than the local escape speed \citep[e.g.,][]{Begelman83,Woods96}. It has been shown that thermal winds arise at 0.1--0.2 $R_C$, the Compton radius. $R_C$ is defined to be $R_C\simeq10^{10} (M/M_\odot)T_{C8}^{-1}$ cm, where $T_{C8}$ is the Compton temperature in terms of 10$^8$ K \citep{Woods96}. Therefore, to launch a thermally driven wind, we require the launching radius, $R_{\rm launch}$, be located at greater than 0.1 $R_C$. If we then assume that the observed velocity is equal to the local escape velocity, we can solve for the corresponding radius, 
\begin{align}
R'_{\rm launch} \simeq & \frac{2G M}{v_{\rm out}^2} \\
= & 10^{11} \left(\frac{M}{M_\odot}\right) \left( \frac{v_{\rm out}}{300 km \ s^{-1}}\right)^{-2} cm .
\end{align}
Setting this radius to be greater than or equal to the launching radius, which is required if the wind is to be thermally driven, we find,
\begin{align}
R'_{\rm launch} & \gtrsim R_{\rm launch} \\
10 \left(\frac{M}{M_\odot}\right) \left( \frac{v_{\rm out}}{300 km \ s^{-1}}\right)^{-2} & > 0.1 \left(\frac{M}{M_\odot} \right) \frac{1}{T_{C8}} \\
T_{C8} > & 10^{-2} \left( \frac{v_{\rm out}}{300 km \ s^{-1}}\right)^{2}.
\end{align}

We can see here that for typical velocities, and low Compton temperature ($<10^6$ K), driving winds by thermal pressure is difficult. We note that the actual velocity is likely to be greater than the line-of-sight velocity due to inclination effects and transverse velocities across our line-of-sight. If the observed velocity is proportional to the gravitational potential. i.e., $v_{\rm out}^2 \gtrsim GM/R_{\rm launch}$, then a higher velocity would place the gas deeper in the potential well and thus increase the temperature needed to launch a thermally driven wind. In addition, we have assumed that the launching radius is the radius at which the observed velocity equals the escape velocity. 

Requiring that the velocity exceeds the escape velocity also requires that the bolometric luminosity to be 
\begin{equation}
L_{\rm Bol} \gtrsim (6.4)^{-3/4} (R/R_{C})^{-1/2} L_{CR} 
\end{equation}
where $L_{CR}$ is the critical luminosity defined as $L_{CR} = 2.88 \times 10^{-2} T_{C8}^{-1/2} L_{\rm Edd}$. \citep[See][for more details.]{Proga00} If the source luminosity is $L_{\rm Bol} \lesssim 2\times 10^{-2} L_{\rm Edd}$ then it would fail to launch an escaping wind. As seen in Figure~\ref{fig:Edd}, a majority of the wind sources are above this threshold, so thermal driving is plausible as long as our assumption about the launching radius is correct. If the wind originates closer than 0.1$R_C$, then other mechanisms are needed.

\begin{figure}
\centering
\includegraphics[scale=.5]{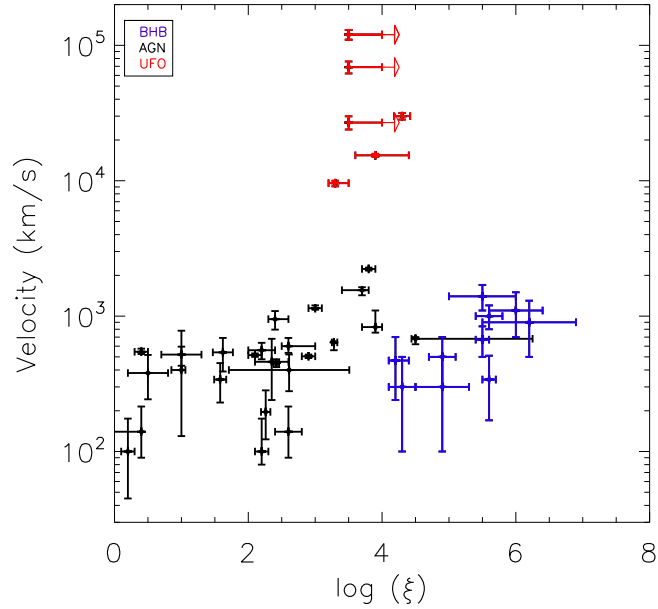}
\figcaption[t]{The above figure plots the observed velocity components versus ionization, for the slow and ``ultra-fast" winds in our black hole sample. In black are the AGN winds, in blue are the BHB winds and in red are the ultra-fast winds. The points with arrows denote lower limits to the ionization state, as the actual state for these ultra fast winds was not analyzed with a photoionization model. \label{fig:vel}}
\end{figure}

\begin{figure}
\centering
\includegraphics[scale=.5,angle=0]{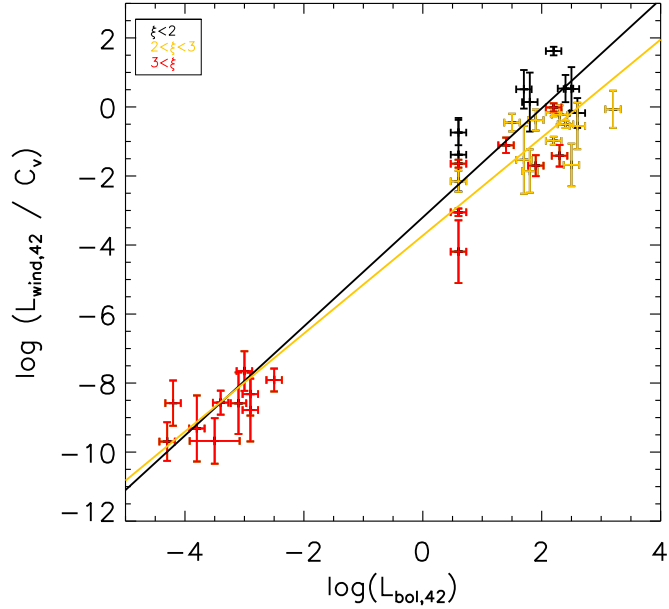}
\figcaption[t]{The plot above shows the correlation between bolometric luminosity and kinetic wind luminosity in individual outflowing components. The line represents the best fit to the total kinetic luminosities which are plotted in Figure~\ref{fig:wind}, while the yellow line is the best fit to the individual components with $\log \xi>2$. The high ionization parameters are described by the following form $\log(L_{\rm wind,42}) = (1.42 \pm 0.06) \log(L_{\rm Bol,42}) - (3.73\pm0.14)$, with an intrinsic scatter of $\sigma_0^{\log \xi>2} =0.57$ \label{fig:indiv}}
\end{figure}

\begin{figure*}
\centering
\includegraphics[scale=.5,angle=90]{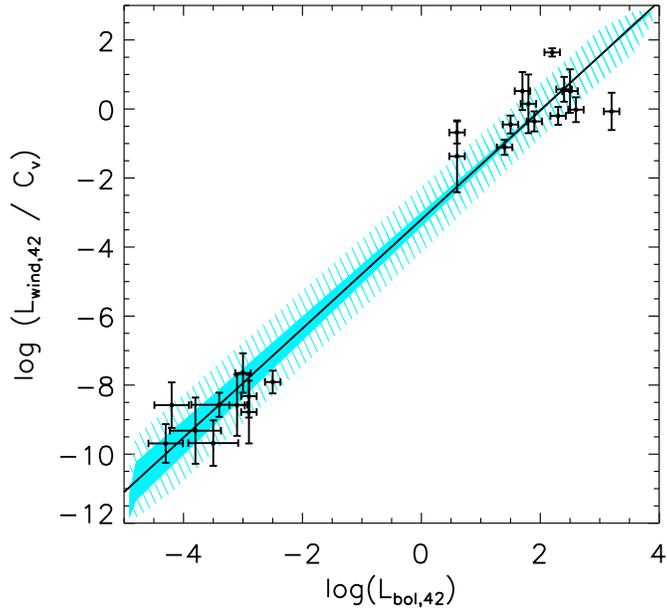}
\figcaption[t]{The plot above shows the correlation between bolometric luminosity and kinetic wind luminosity. The black line is described by $\log (L_{\rm wind,42}) = (1.58 \pm 0.07) \log (L_{\rm Bol,42}) -(3.19\pm0.19)$, with an intrinsic scatter of $\sigma_0 =0.68$. The blue dashed region is the 1$\sigma$ confidence region including the scatter of the relation. The solid region is the 1$\sigma$ confidence region excluding the scatter. The wind kinetic luminosity is plotted per filling factor. The plot shows a simple regulation of wind production across a large mass scale, and the slope indicates that the SMBH winds are more efficient then the stellar-mass black holes. \label{fig:wind}}
\end{figure*}

\begin{figure*}
\centering
\includegraphics[scale=.5,angle=90]{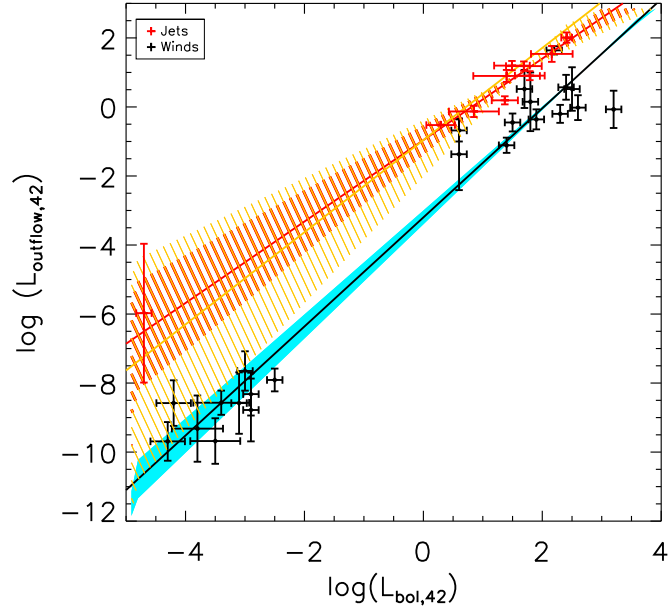}
\figcaption[t]{The plot above depicts the wind power versus the bolometric luminosity, just like Figure~\ref{fig:wind}, but this figure includes the jet power as red data points. The red line describes all the jet points as $\log (L_{\rm Jet,42}) = (1.18 \pm 0.24) \log (L_{\rm Bondi,42}) -(0.96\pm0.43)$. The yellow line describes the data set if Cygnus X-1 is excluded from the fit is given as $\log (L_{\rm Jet,42}) = (1.34 \pm 0.50) \log (L_{\rm Bondi,42}) -(0.80\pm0.82)$. The dashed regions are the 1$\sigma$ confidence regions. The orange line and dashed region is the best fit line and 1$\sigma$ confidence region when excluding Cygnus X-1 from the fit. One can see that the normalization of the jets is higher, demonstrating that for a given bolometric luminosity they are more powerful. One can also see that the slope between the two relations is quite similar, perhaps indicating a common launching mechanism\label{fig:jet}}
\end{figure*}

\begin{figure*}
\centering
\includegraphics[scale=.5,angle=0]{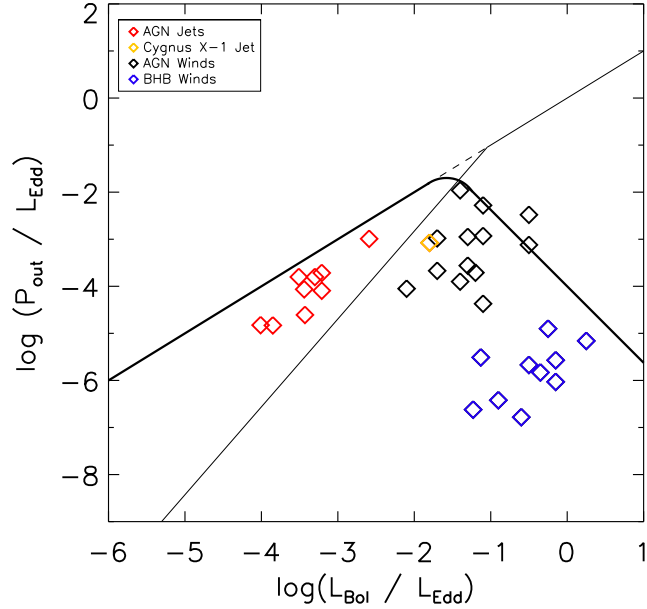}
\figcaption[t]{The above plot shows the power emitted either from the jet power (AGN:red and BHB:orange) or wind power (AGN:black and BHB:blue) as a compared to the mass accretion rate, which is approximated by the bolometric luminosity on the x axis. A clear turnover at $\dot{M}_{acc} \approx 10^{-2} \dot{M}_{\rm Edd}$ indicates where the power emitted is becoming less efficient. Interesting is the dichotomy between where the jets lie at lower mass accretion rates and where the winds lie at higher accretion rates. The thick black line denotes the output power by outflows, where as the thin line is the power generated by radiation as described by \cite{Churazov05} \label{fig:Edd}}
\end{figure*}

\begin{figure*}
\centering
\includegraphics[scale=.5,angle=90]{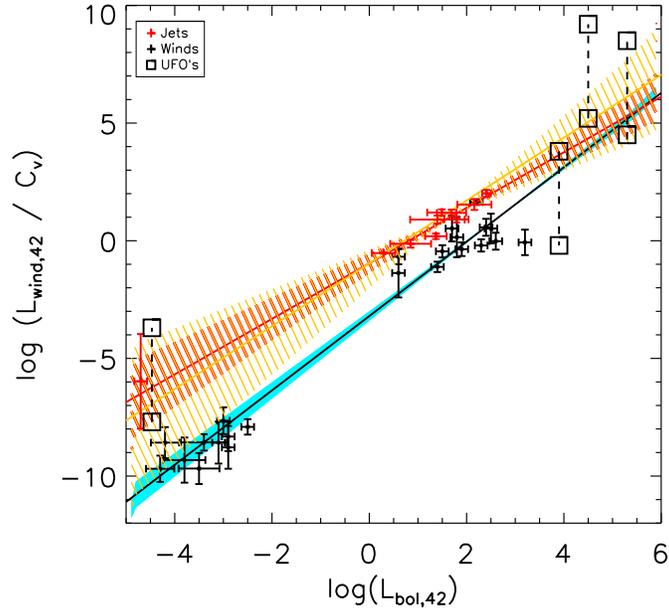}
\figcaption[t]{This plot is the same as Figure~\ref{fig:jet} but now includes UFO's in black squares ($v>0.1c$). The upper square is the power estimate with a filling factor of unity. The bottom square connected by the dashed line is the lower estimate of the wind power if the filling factor is as low as $C_v=10^{-4}$. Even with a smaller filling factor, the UFO's resemble the jet relation more so then the wind relation. Perhaps this is indicated that the winds are reaching a phase where they are being accelerated into jets.\label{fig:ufo}}
\end{figure*}

\begin{figure*}
\centering
\includegraphics[scale=.5,angle=90]{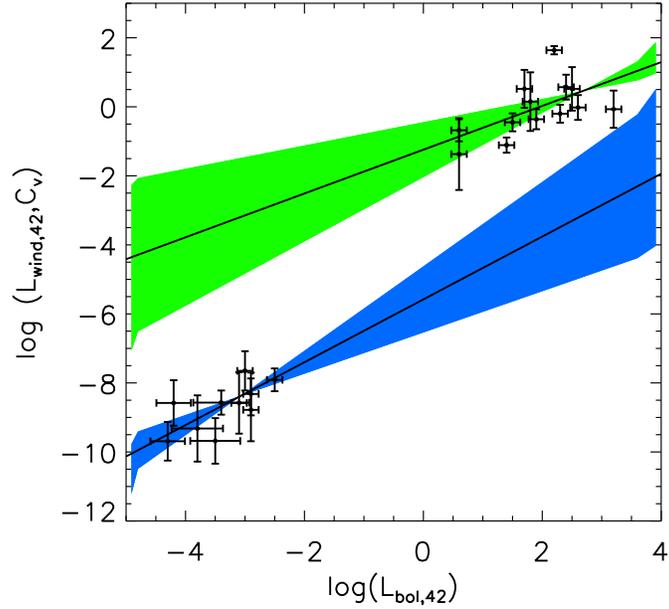}
\figcaption[t]{This figure shows the best fit linear regressions  when the AGN and BHB samples are fit separately. The BHB are described by $\log (L_{\rm wind,42}^{\rm BHB}) = (0.91 \pm 0.31) \log (L_{\rm Bol,42}^{\rm BHB}) -(5.58\pm 1.68)$ with scatter consistent with zero. The AGN are described by $\log (L_{\rm wind,42}^{\rm AGN}) = (0.63 \pm 0.30) \log (L_{\rm Bol,42}^{\rm AGN}) -(1.24\pm0.63)$ with scatter $\sigma^{\rm AGN}_0 = 0.58$.\label{fig:indivmass}}
\end{figure*}

\begin{figure*}
\centering
\includegraphics[scale=.5,angle=90]{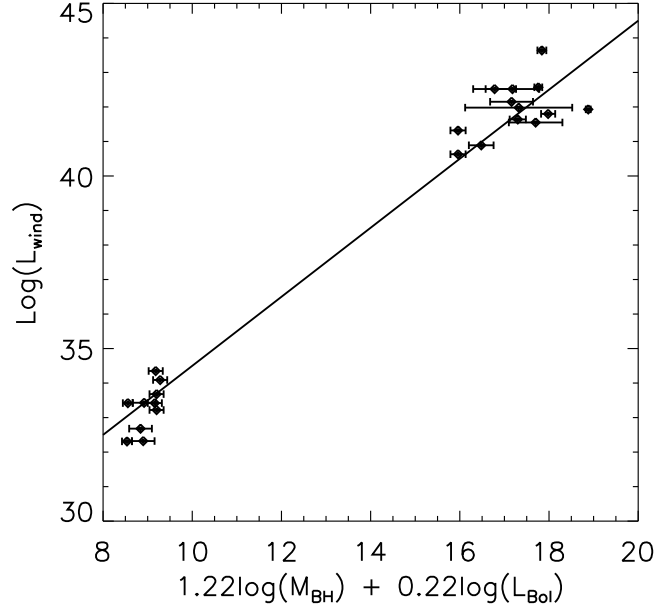}
\figcaption[t]{The above plot shows the best fit plane of our wind sample when including mass as a third parameter with bolometric luminosity and wind power. The plane is described by $\log (L_{\rm wind} ) = (1.2\pm 0.3) \log(\rm M_{BH}) + (0.2 \pm 0.4) \log (L_{\rm Bol}) + (24.5 \pm 0.2)$ with scatter $\sigma_0=0.68$.\label{fig:fp}}
\end{figure*}

\begin{figure*}
\centering
\includegraphics[scale=.5,angle=90]{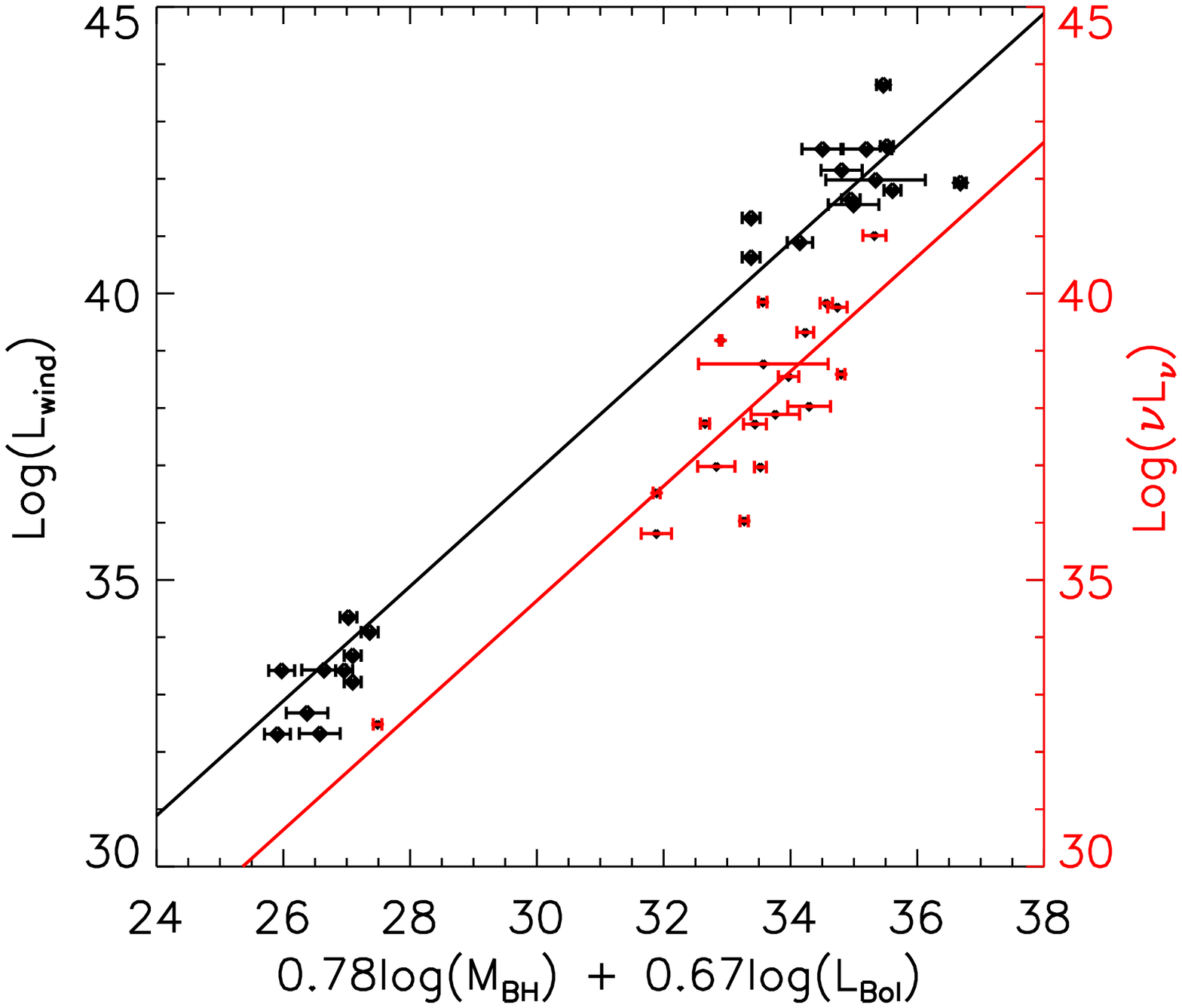}
\figcaption[t]{This plot shows our wind data plotted against the fundamental plane of black hole activity described by \cite{Gultekin09}. The solid line is used to show the one-to-one correspondence in this plane cross section. $\nu$=5GHz. Although the intercepts are different, the coefficients of mass and X-ray/Bolometric luminosity are consistent between our sample and \cite{Gultekin09}, which may tentatively suggest a common driving mechanism. \label{fig:Gultekin}}
\end{figure*}

\cite{Luketic10} perform hydrodynamical simulations to explore whether thermal driving could be responsible for the winds seen in X-ray binaries. They conclude that at low densities, thermal driving is possible from an X-ray heated accretion disk. However, at densities higher than $n_e > 10^{12}$ cm$^{-3}$, Compton heating is not sufficient at driving winds at velocities of $v_{\rm out} \geq 10^{2}$ km s$^{-1}$. \cite{Luketic10} compare their work to observations of GRO J1655-40 which has a high density of $n_e \simeq 10^{14}$ cm$^{-3}$ \citep{Miller08}, and conclude that thermal driving is not responsible for its winds. It is possible that the other X-ray binaries have similar densities, and as Figure~\ref{fig:vel} shows, they have similar velocities as well as high ionizations. Therefore, Compton heating may be an unlikely driving source for these X-ray binaries. 

The AGN in Figure~\ref{fig:vel} show a much wider range in ionization but are all outflowing at velocities consistent with $v_{\rm out}>10^{2}$ km s$^{-1}$. If they share a similar density to that of the BHB, then Compton heating is not a viable driving mechanism for them either. \cite{Dorodnitsyn08} also perform hydrodynamical simulations for Compton heated and radiation driven AGN winds. They find similar outflowing velocity and ionization parameters as we show in Figure~\ref{fig:vel} but the density they assume is far lower than what is inferred from observations \citep[e.g.,][]{Mckernan07,Risaliti09,King12b}. In addition, the location of their warm-absorbers are much farther from the central source than those inferred from observations \citep[e.g.,][]{King12b,Crenshaw12}.

A third driving mechanism can be magnetic fields, whether through magneto-centrifugal force \citep{Blandford82}, or magnetic pressure from the toroidal field generated by MRI in the disk, as suggested by \cite{Contopoulos95}, \cite{Miller00}, and \cite{Proga03}. These winds tap magnetic field energy generated or sustained in the disk. It has been shown that for at least three of the sources included in our study, GRO 1655-40, NGC 4051, and NGC 4151, magnetic processes are likely driving the observed winds \citep{Kraemer05,Miller08,King12b,Neilsen12}. Because these sources (the BHB, GRO 1655-40, and Seyfert-1's, NGC 4051 and NGC 4151) are included in this relation, and span orders of magnitude in mass, it raises the possibility that magnetic forces may drive this wind relation. 

Jets are also thought to be driven by magnetic processes in the disk or near the black hole \citep[e.g.,][]{Lovelace76,Blandford82, Blandford77,Krolik10}. The jet power relation is determined to be $\log (L_{\rm Jet,42})= (1.18 \pm 0.24) \log (L_{\rm Bol,42})- (0.96 \pm 0.43)$. If we exclude Cygnus X-1, then the jet relation becomes $\log (L_{\rm Jet,42})= (1.34 \pm 0.50) \log (L_{\rm Bol,42}) -(0.80 \pm 0.82)$ (See Figure~\ref{fig:jet}). When we include Cygnus X-1 in the jet relation the slope of the jet relation and the initial wind relation are inconsistent at the 1.6$\sigma$ level. When we exclude Cygnus X-1, the jet relation slope is formally consistent with that of the single wind relation. Further, if we examine the individual wind fits, the jet relation including or excluding Cygnus X-1 is consistent with the BHB slope. When comparing the jet relation to the AGN sample, the slopes are only inconsistent at the 1.4$\sigma$ and 1.2$\sigma$ level, including and excluding Cygnus X-1, respectively. Finally, we very tentatively suggest that the plane fit to the wind sample is consistent with the fundamental plane of accretion onto black holes \citep{Gultekin09}, which would be further evidence of the similar dependence on mass accretion rate (as well as mass) of both the winds and jets. 

If the two types of outflows are regulated by the mass accretion rate in the same fashion, then the same driving mechanism may also be at work, as the geometry or mass loading of the magnetic fields may be driven by the mass accretion rate as well. This may explain the formal consistency of the slopes between the jet and wind relations. A \cite{Blandford82} scenario may be a viable solution for driving these outflows, and could be possibly aided by \cite{Blandford77} scenario for jets. In addition, \cite{Ohsuga11} show that MHD accretion flows can drive both jets and winds depending on the mass accretion rate, qualitatively consistent with Figure~\ref{fig:Edd}.

The ultra-fast outflows appear to follow the jet relation (See Figure~\ref{fig:ufo}). These are winds whose observed velocity exceeds $v>0.01c$. This raises the question of how these ultra-fast winds are accelerated to such high velocities. Are we seeing the phase at which these winds are being collimated into jets, as the power associated with the winds is very comparable to the jet power? Again, this could point to a shared driving mechanism between winds and jets, such as MHD \citep{Lovelace76,Blandford82}, if we are truly observing this transition phase between the two. 

\subsection{Implications for Feedback}
\label{feedback}
\begin{figure*}
\centering
\includegraphics[scale=.5,angle=90]{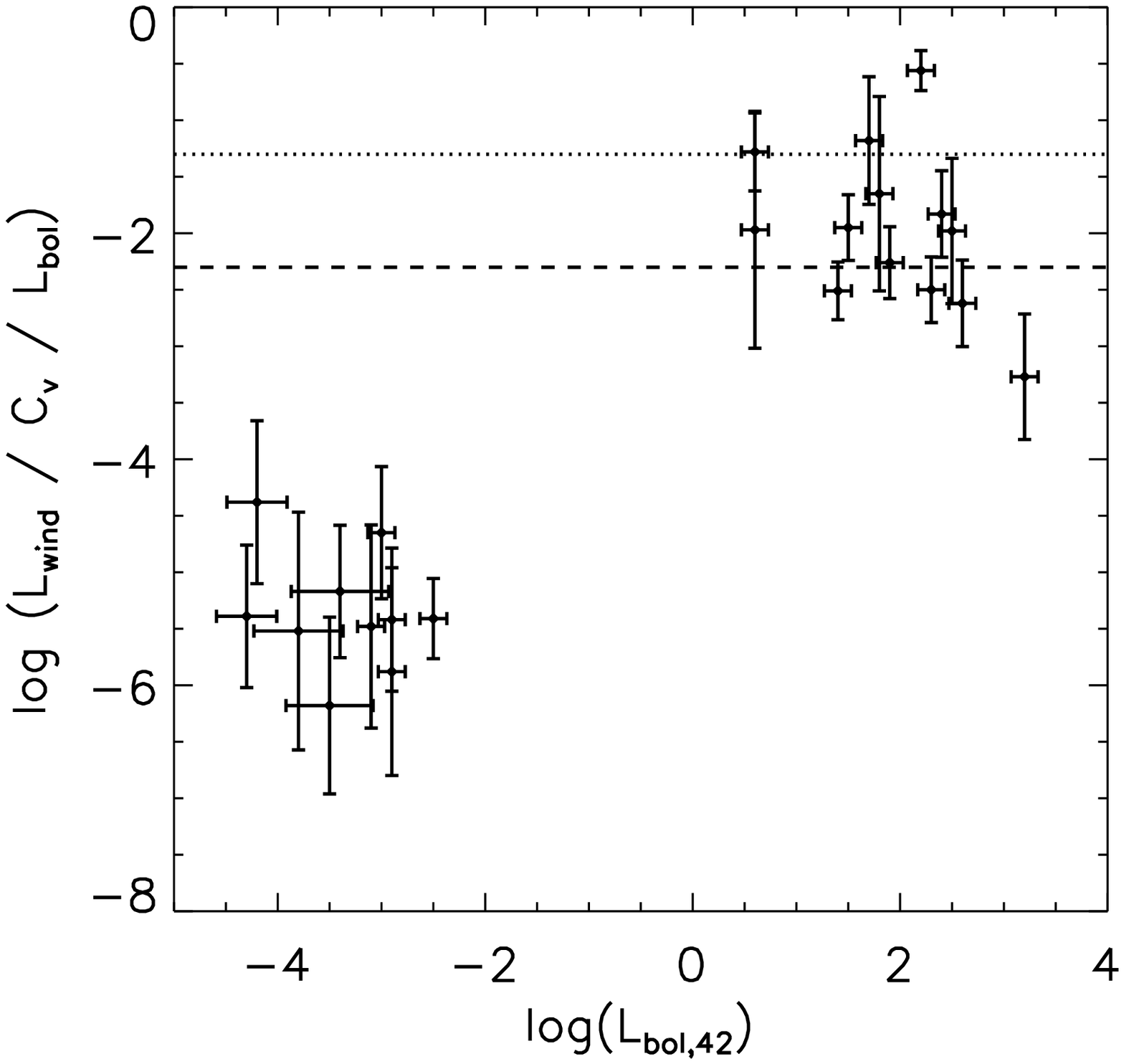}
%\centerline{~\psfig{file=/Atomic1/ashking/windsVjets/wind.ps,width=3.4in,angle=-90}~}
\figcaption[t]{This plot shows the correlation between kinetic wind luminosity per filling factor divided by the bolometric luminosity as compared to the bolometric luminosity. The dotted line is 5\% $L_{\rm Bol}$, while the dashed line is 0.5\% $L_{\rm Bol}$. These are the limits of the kinetic wind power reported by \cite{Dimatteo05} and \cite{Hopkins10}, respectively, for mechanical feedback to have an influence on black hole growth and feedback. We expect these winds to have a small filling factor which would make the wind power estimate lower, and perhaps below 0.5\% $L_{\rm Bol}$. \label{fig:bol}}
\end{figure*}

The characterization of these outflows allows us to determine that X-ray AGN winds are more efficient at removing material than are X-ray BHB winds. Interestingly, \cite{Hopkins10} show that only 0.5\% of the bolometric luminosity needs to be converted into mechanical power in order to regulate black hole growth and affect feedback in the host galaxy. As shown in Figure~\ref{fig:bol}, the majority of the AGN lie above (or are consistent with) $5\times10^{-3}L_{\rm Bol}$. A few sources lie above $5\times10^{-2}L_{\rm Bol}$ (the dotted line). \cite{Crenshaw12}, in a study focusing only on AGN winds, show that up to half of their AGN are consistent with $\gtrsim 5\times10^{-3} L_{\rm Bol}$. However, if the filling factor is much less than unity, the wind power will be far less than the $5\times10^{-3}L_{\rm Bol}$ limit for influential feedback. This may imply that the X-ray winds do not have a large impact on feedback.

In addition, Figure~\ref{fig:indiv} shows a stratification of the kinetic wind luminosity as a function of ionization in the AGN. The low ionization components ($\log \xi<2$) tend to have a much higher kinetic luminosity as compared to the medium ionization components ($2<\log \xi <3$) and high ionization components ($3<\log\xi$). The reason for this may again be because the filling factor is not included in this analysis. As mentioned in Section~\ref{methods}, these black hole X-ray winds are thought to be clumpy and filamentary. Moreover, observations of ionized stellar winds indicate that the less ionized gas should be more clumpy, i.e. have a lower filling factor, due to pressure confinement from the hot surrounding gas \citep[e.g.,][]{Sako99}. This would imply that the low ionization components seen in Figure~\ref{fig:indiv} would likely have a lower filling factor than the high ionization components. If the filling factors were included, the low ionization components would no longer rise above the higher ionization components. This would serve to flatten the initial wind relation, making the wind slope even more consistent with the jet relation slope. Feedback from the lowest ionization components would no longer dominate the relation. 

On the other hand, the low ionization components may also be consistent with being radiatively driven, and therefore would not follow the same relation as the high ionization components anyway. As shown in Section \ref{bol}, the high ionization components do follow a shallower slope of $\alpha^{\log \xi>2} = 1.42 \pm 0.06$, which is consistent with both jet relations, i.e. including and excluding Cygnus X-1. 

Regardless of the ionization of the winds, jets are more efficient at a given bolometric luminosity, compared to X-ray winds. When considering power alone, jets may have a greater impact on mechanical feedback and galaxy evolution then winds. Depending on the mass accretion rate, for which we use the bolometric or Bondi luminosity as a proxy, we can now characterize the associated jet and wind power. Figure~\ref{fig:Edd} shows exactly how the outflow power scales with Eddington fraction. There is a division between dominant outflow at approximately $10^{-2} L_{\rm Edd}$. If both the winds and jets share a common launching mechanism, this division may be strongly driven by the mass accretion rate. Mass loading or even the geometry of the magnetic fields in the disk would have an important role as well, and can again be directly regulated by the mass accretion rate. The transition seen at approximately $10^{-2} L_{\rm Edd}$ is also interesting because this is the regime where winds begin to prevail over jet production, especially seen the spectral state dependence in X-ray binaries. 

We can now describe the outflow power as a function of Eddington fraction directly associated with the inner-accretion disk surrounding a black hole, vital for cosmic simulations. This is important because as Figure~\ref{fig:Edd} demonstrates, outflows are present in a range of Eddington fractions, not just low Eddington fractions. As galaxies evolve through their ``Quasar" and ``radio" modes of accretion, we are still able to prescribe the outflowing power to assess the mechanical feedback in those systems and explore the implications for galactic co-evolution.

\subsection{Potential Caveats}
\label{Caveats}
Before using these descriptions, it is important to understand the caveats involved in assembling this data set. As shown in Figure \ref{fig:Edd}, there is a potential state dependence of outflow type on accretion rate \citep[also see][]{Miller06b,Miller08,Neilsen09,Ponti12}. However, it is not clear if this is a result of a selection bias toward high luminosity AGN. One could imagine that at low X-ray luminosity, i.e. ellipticals and BHB in the ``low/hard" state, detections of winds could be hampered by low signal-to-noise. This would be most pertinent to our jet sample, which is dominated by low luminosity AGN accreting at low accretion rates. However, even if winds were to coexist in these low accretion rates, just as they do in Seyfert 1 NGC 4051 \citep{King11}, jet power is likely to dominate by orders of magnitude, as the wind power is proportional to the ionizing luminosity, which would be small. In addition, in BHB strong limits to wind detection have been made in the ``low/hard" state \citep[e.g.,][]{Neilsen09} as well as strong upper limits to jet production in the ``high/soft" state \citep[e.g.,][]{King12a}. Therefore, we stress that this work is focused on the dominant outflow. 

We also note the difficulty in placing upper limits on wind detections using absorption features in the X-ray band. As these features can be seen at an array of different velocities and ionization states, there is no specific wavelength one would expect to find an absorption feature denoting an outflow. Further, the wind power estimates do not depend on the strength of the line, but only the wavelength and ionization state, making estimates of upper limits rather difficult. These issues of detection affect both BHB and AGN in the same manner, and we stress that the lack of detection of these absorption features is not evidence for the absence of a wind, but may be the absence of evidence. Again, the state dependence of outflows seen in BHB and now in AGN (Figure \ref{fig:Edd}) is likely to be driven by accretion rate. However, longer integrations to improve signal-to-noise of BHB in the ``low/hard" state and AGN at low accretion rates are needed to be confident of this assessment \citep[e.g.,][]{Miller12}.
Next, we note that there are outflows other than the ones examined in our analysis that are still important in removing substantial amounts of material from their accretion disks and host galaxies. In particular, broad absorption line (BAL) quasars have particularly powerful outflows \citep{Moe09,Dunn10,Brandt00}. However, these outflows are observed in the optical and ultra-violet regime and have much lower ionization parameters than the X-ray winds discussed here. Therefore, they are not as readily associated with - or driven by - the inner accretion disk, and have not been included in our analysis. \cite{Crenshaw12} also show a positive correlation between ionization parameter, $U$, and column density in local AGN in their figure 3. This demonstrates that the bulk of the outflow material is being observed in the X-ray regime.

A broader range of ionization parameters are probed in the AGN as compared to the BHB (See Figure~\ref{fig:vel}), which may also contribute to the AGN scatter. The mix of $\xi$ in AGN calls into question whether we are probing the same physics, i.e. closest to the black holes. Because ionization is dependent on the distance as $\xi = L (nr^2)^{-1}$, similar ionization states should probe the same distance from the black hole for a given luminosity. Consequently, similar micro-physics at a given radius and ionization should be at work. Further, in Section \ref{feedback}, examining the high ionization components alone results in a shallower slope when comparing bolometric luminosity to wind power. A shallower relation is more consistent with the jet relation. This demonstrates the clear need for a much larger sample size. Fortunately, {\it Astro-H} will provide the needed coverage in the highest ionization band. This will not only allow for the detection of additional sources, but also detection of the highest ionization states for comparison with stellar-mass black holes.

Although we see a large range in ionization, we do not see as large of a range in velocity. It is only when we include the ultra-fast outflows that three orders of magnitude in velocity are probed as compared to the six orders of magnitude in the ionization parameter. This trend is important in understanding whether inclination has an effect on the given $L_{\rm wind}$ vs $L_{\rm Bol}$ correlation. As BHB winds are thought to be observed in nearly edge-on sources \citep[e.g.][]{Miller06b,Miller06,Ponti12}, and AGN winds, especially Seyfert 1 AGN winds, are thought to be observed in face-on sources \citep[e.g.,][]{Wu01}, inclination could have the potential to bias our results. However, the data show no trend in velocity as a function of inclination. In addition, when examining BHB sources individually, face-on sources (e.g., GX 339--4, XTE J1817--330) do not show absorption features in the Fe K band when they are in the ``high/soft" state, contrary to their edge-on counterparts. This is due to limited sensitivity, since low inclination sources tend to be softer and to give less signal through the Fe K band. As noted in \cite{Ponti12}, the limits on flux in face-on sources are not very constraining, and lines as weak as those in H1743-322 \citep{Miller06} could not have detected in e.g. XTE J1817-330. For instance, there is likely a simple absence of evidence for BHB winds in face-on systems. 

One may also expect inclination to have an effect on the estimated kinetic jet power. For example, those sources for which the jet is directed along our line of sight may suffer from Doppler boosting. However, this would primarily affect the radio luminosity of such sources, and not the kinetic power, which is taken from estimates of cavity sizes. On the other hand, the jet power may be influenced by the spin of the black hole. If jets are driven by the \cite{Blandford77} mechanism, then spin may play a large role in the power released by the jets. However, the common slopes between the jet and wind relation points to more of a \cite{Blandford82} scenario, where the spin of the black hole does not affect the power released. Further, the fact that the jet and wind power seem to be present at certain Eddington fractions, point to the idea that mass accretion rate may be the throttle that is ultimately driving the type and power of the outflow. We note that spin is unlikely to play a large role in the X-ray wind power regardless, as winds are thought to originate further out in the accretion disk.

An additional concern with the jet power is that the estimate is a long-term average and not instantaneous as are the wind power estimates. If the jet production occurs on timescales that are much shorter than the dynamical timescale of the cavity, than the power estimates would increase. Unfortunately, this is a limit of this technique when using cavities to estimate power. However, long term estimates of power are more pertinent for feedback estimates. 

Finally, as this sample is small in size, the results must be regarded cautiously and tested in the future. It is imperative that we obtain more observations at all masses and mass accretion rates. Specifically, black holes accreting at $L_{Bol} \sim10^{41}-10^{42}$ ergs s$^{-1}$ could distinguish whether one linear fit is required across the entire wind sample or if the BHB and AGN are better fit by individual linear fits. This could either be a small Seyfert galaxy with mass on order of $M\sim 10^5 M_\odot$ accreting at a few percent of Eddington, or a large SMBH, $M\sim10^9 M_\odot$, accreting at a very low Eddington rate. In addition, non-simultaneity of AGN luminosities could have a dramatic effect on the observed scatter seen in the X-ray winds. Although the AGN timescales for disk evolution are longer than BHB, observations made years apart may not probe the same accretion regime. 

\section{Conclusions}
\begin{itemize}
\item In this study, we find that winds are consistent with being regulated according to a simple relation across a large mass scale. In particular, we find the trend is described as $\log L_{\rm wind,42} \propto (1.58 \pm 0.07) \log L_{\rm Bol,42}$. The slope is greater than unity, so it may imply that the SMBH are more efficient at expelling material than BHBs. 
\item If we fit the BHB and AGN populations separately, they still require consistent slopes of $\alpha^{BHB} = 0.91\pm0.31$ and $\alpha^{AGN} = 0.63\pm0.30$. Further, if we assume mass is influencing this relation and fit a plane to the data, we find the best fit relation to be $\log (L_{\rm wind} ) = (1.2\pm 0.3) \log(\rm M_{BH}) + (0.2 \pm 0.4) \log (L_{\rm Bol}) + (24.5 \pm 0.2)$ with scatter $\sigma_0=0.68$ consistent with the ``fundamental plane" of accretion onto black holes.
\item It remains possible that different processes tied to the mass accretion rate- thermal driving in stellar-mass black holes and radiative driving in AGN - are actually at work in driving winds. However, it is not clear that these different mechanisms should agree so well and follow the same slope in these wind relations. Moreover, it seems that a magnetic wind must be at work in GRO 1655$-$40, NGC 4051, and NGC 4151 \citep{Miller08,King12b,Kraemer05}, which fall on the relation. This may also suggest a role for magnetic driving across the mass scale.
\item Furthermore, when we examine jet power, the data may be consistent with winds and jets being regulated in a common fashion. Since radiative and thermal processes are not likely to drive relativistic jets, a mechanism like magnetocentrifugal or MHD winds are plausible explanations \citep[e.g.,][]{Blandford82,Proga03}. 
\item The ultra-fast winds appear to obey the same regulation scheme as slower, more common winds, if they have a low filling factor and the slow winds have a high filling factor close to unity. However, some ultra-fast winds appear to carry as much kinetic luminosity as jets, even after accounting for filling factors. This suggests that we may be seeing a phase where winds finally are accelerated into jets. 
\item Figure \ref{fig:Edd} provides a direct way to quantify the outflow power as a function of mass accretion rate. A division between dominant outflow state is observed at approximately $10^{-2} L_{\rm Edd}$. This trend has broad implications, especially for theoretical simulations that need prescriptions for feedback to study galactic dynamics and evolution. 
\item A larger sample will help us to distinguish between these proposed relations as well as quantify the intrinsic scatter. As it stands now, {\it Chandra} will play an integral part in future studies. Looking further ahead, {\it Astro-H} will have improved sensitivity in the Fe K band, enabling unprecedented looks at the most ionized and innermost flows in the accretion disks of both BHB and AGN. 
\end{itemize}
\section*{Acknowledgements}
The authors would like to thank the anonymous referee for their invaluable comments to improve this paper. ALK acknowledges support from NASA through the NESSF program. JMM thanks NASA for support through its guest observer programs.
\bibliography{bibwinds}
\clearpage
%-------------------------------------Table Start---------------------------------------------------------
\begin{deluxetable}{l l l l l l l}
\tablecolumns{7}
\tablewidth{0pc}
\tabletypesize{\scriptsize}
\tablecaption{Individual X-ray Wind Components}
\tablehead{ Object & Type & Component &$\log \xi$ & Velocity & $L_{\rm wind}/C_v$ & Reference\\
& & & ergs cm s$^{-1}$ & (km s$^{-1})$ &(ergs s$^{-1}$) & }
\startdata
{\bf SMBH} \\
\hline Akn 564 & S1 &
1 & 0.40 $\pm$ 0.25 & 140 $\pm$ 62 & 42.52 $\pm$ 0.63 & \cite{Mckernan07} \\
& & 2 & 2.60 $\pm$ 0.20 & 140 $\pm$ 62 & 40.32 $\pm$ 0.62 \\
\hline IC 4329a & S1 &
1 & 0.20 $\pm$ 0.10 & 100 $\pm$ 65 & 42.14 $\pm$ 0.85 & \cite{Mckernan07}\\
& & 2 & 2.20 $\pm$ 0.10 & 100 $\pm$ 47 & 40.14 $\pm$ 0.63 \\
\hline IRAS 18325 & S2 &
1 & 1.58 $\pm$ 0.09 & 340 $\pm$ 110 & 41.82 $\pm$ 0.43 & \cite{Zhang11b}\\
& & 2 & 2.35 $\pm$ 0.25 & 460 $\pm$ 220 & 41.45 $\pm$ 0.67 \\
\hline MCG -6-30-15 & S1 &
1 & 3.70 $\pm$ 0.20 & 1555 $\pm$ 105 & 40.89 $\pm$ 0.22 & \cite{Mckernan07}\\
\hline Mrk 290 & S1 &
1 & 1.62 $\pm$ 0.15 & 540 $\pm$ 150 & 42.53 $\pm$ 0.39 &\cite{Zhang11}\\
& & 2 & 2.42 $\pm$ 0.04 & 450 $\pm$ 30 & 41.50 $\pm$ 0.11 \\
\hline Mrk 509 & S1 &
1 & 2.26 $\pm$ 0.07 & 196 $\pm$ 80 & 41.93 $\pm$ 0.54 & \cite{Ebrero11}\\
\hline NGC 3516 & S1 &
1 & 2.40 $\pm$ 0.15 & 950 $\pm$ 147 & 41.55 $\pm$ 0.26 &\cite{Mckernan07} \\
\hline NGC 3783 & S1 &
1 & 2.90 $\pm$ 0.10 & 505 $\pm$ 15 & 41.02 $\pm$ 0.12 &\cite{Mckernan07} \\
& &2 & 2.10 $\pm$ 0.10 & 515 $\pm$ 15 & 41.85 $\pm$ 0.12 \\
& &3 & 0.40 $\pm$ 0.10 & 545 $\pm$ 25 & 43.62 $\pm$ 0.12 \\
& & 4 & 3.00 $\pm$ 0.10 & 1145 $\pm$ 42 & 41.99 $\pm$ 0.12 \\
\hline NGC 4051 & S1 &
1 & 1.00 $\pm$ 0.30 & 520 $\pm$ 82 & 41.26 $\pm$ 0.37 & \cite{Mckernan07}\\
& & 2 & 2.60 $\pm$ 0.25 & 600 $\pm$ 77 & 39.85 $\pm$ 0.30 \\
& & 3 & 3.80 $\pm$ 0.10 & 2230 $\pm$ 55 & 40.36 $\pm$ 0.11 \\
\hline NGC 4051 & S1&
1 & 4.50 $\pm$ 0.90 & 680 $\pm$ 40 & 37.81 $\pm$ 0.91 &\cite{King12a}\\
& & 2 & 3.28 $\pm$ 0.04 & 640 $\pm$ 45 & 38.95 $\pm$ 0.11 \\
& & 3 & 1.00 $\pm$ 0.11 & 400 $\pm$ 325 & 40.62 $\pm$ 1.06 \\
\hline NGC 4151 & S1 &
1 & 3.58 $\pm$ 0.30 & 491 $\pm$ 8 & 40.30 $\pm$ 0.30 &\cite{Kraemer05} \\
& & 2 & 2.26 $\pm$ 0.30 & 491 $\pm$ 8 & 41.62 $\pm$ 0.30 \\
\hline NGC 4593 & S1 &
1 & 2.61 $\pm$ 0.90 & 400 $\pm$ 121 & 40.47 $\pm$ 0.98 &\cite{Mckernan07} \\
& & 2 & 0.50 $\pm$ 0.30 & 380 $\pm$ 137 & 42.51 $\pm$ 0.56 \\
\hline NGC 5548 & S1 &
1 & 2.20 $\pm$ 0.20 & 560 $\pm$ 77 & 41.77 $\pm$ 0.27 & \cite{Mckernan07}\\
& & 2 & 3.90 $\pm$ 0.15 & 830 $\pm$ 172 & 40.59 $\pm$ 0.31 \\
\hline {\bf BHB} \\
\hline 4U 1630 & &
1 & 4.90 $\pm$ 0.40 & 300 $\pm$ 200 & 32.68 $\pm$ 0.96 & this paper\\
\hline GRO 1655$-$40 & &
1 & 4.90 $\pm$ 0.20 & 500 $\pm$ 200 & 32.31 $\pm$ 0.56 & \cite{Miller08}\\
\hline GRO 1655$-$40 & &
1 & 4.20 $\pm$ 0.15 & 470 $\pm$ 230 & 33.42 $\pm$ 0.66 & \cite{Neilsen12} \\
\hline H 1743$-$322 a & &
1 & 5.50 $\pm$ 0.10 & 670 $\pm$ 170 & 33.43 $\pm$ 0.35 & this paper \\
\hline H 1743$-$322 b & &
1 & 5.60 $\pm$ 0.10 & 340 $\pm$ 170 & 32.32 $\pm$ 0.66 & this paper \\
\hline GRS 1915$+$105 s1 & &
1 & 4.30 $\pm$ 0.20 & 300 $\pm$ 200 & 33.42 $\pm$ 0.89 & \cite{Ueda09} \\
\hline GRS 1915$+$105 s2 & &
1 & 5.60 $\pm$ 0.20 & 1000 $\pm$ 200 & 34.09 $\pm$ 0.33 & \cite{Miller06}\\
\hline GRS 1915$+$105 s3 & &
1 & 5.50 $\pm$ 0.50 & 1400 $\pm$ 300 & 34.35 $\pm$ 0.57 & this paper\\
\hline GRS 1915$+$105 s4 & &
1 & 6.00 $\pm$ 0.40 & 1100 $\pm$ 400 & 33.68 $\pm$ 0.62 & this paper \\
\hline GRS 1915$+$105 s5 & &
1 & 6.20 $\pm$ 0.70 & 900 $\pm$ 400 & 33.22 $\pm$ 0.91 & this paper\\
\enddata 
\label{tab:indiv}
\tablecomments{The above table lists all the components that are considered in this analysis. S1 stands for Seyfert 1 and S2 stands for Seyfert 2. The kinetic luminosity from the AGN that have more than one component for a single observation are summed and included as total kinetic luminosities in Table~\ref{tab:lum}}
\end{deluxetable}
%-------------------------------------Table End---------------------------------------------------------
%-------------------------------------Table Start---------------------------------------------------------
\begin{deluxetable*}{l l l l l l l l l}
\tablecolumns{9}
\tablewidth{0pc}
\tabletypesize{\scriptsize}
\tablecaption{X-ray Wind and Jet Quantities}
\tablehead{ & & Object & $\log L_{\rm Bol}$ & $\log L_{\rm wind} / C_v$ &$\log M_{BH}$ & $\log D$ & Code & Reference\\
&&& (ergs s$^{-1}$) & (ergs s$^{-1}$) &($M_{\odot}$) & (cm) & & }
\startdata
\startdata
{\bf X-ray winds} \\ \hline
\multirow{12}{*}{AGN} & 1 & Akn 564* & 44.50 $\pm$ 0.13 & 42.52 $\pm$ 0.63 & 6.9$^e$ & 26.51 & XSTAR & \cite{Mckernan07} \\
& 2 & IC 4329a* & 43.80 $\pm$ 0.13 & 42.15 $\pm$ 0.85 & 7.0$^c$ & 26.32 & XSTAR & \cite{Mckernan07} \\
& 3 & IRAS 18325* & 44.60 $\pm$ 0.13 & 41.98 $\pm$ 0.36 & 7.0$^b$ & 26.42 & XSTAR & \cite{Zhang11b}\\
& 4 & NGC 3516 & 43.50 $\pm$ 0.13 & 41.55 $\pm$ 0.26 & 7.5$^a$ & 26.07 & XSTAR & \cite{Mckernan07}\\
& 5 & NGC 3783 & 44.20 $\pm$ 0.13 & 43.64 $\pm$ 0.12 & 7.5$^c$ & 26.11 & XSTAR & \cite{Mckernan07} \\
& 6 & NGC 4051 & 42.60 $\pm$ 0.13 & 41.32 $\pm$ 0.32 & 6.2$^a$ & 25.50 & XSTAR & \cite{Mckernan07} \\
& 7 & NGC 4051 & 42.60 $\pm$ 0.13 & 40.63 $\pm$ 1.04 & 6.2$^a$ & 25.50 & XSTAR & \cite{King12b} \\
& 8 & NGC 4151* & 43.90 $\pm$ 0.13 & 41.64 $\pm$ 0.29 & 7.1$^c$ & 25.64 & XSTAR & \cite{Kraemer05} \\
& 9 & NGC 4593 & 43.70 $\pm$ 0.13 & 42.52 $\pm$ 0.55 & 6.7$^c$ & 26.04 & XSTAR & \cite{Mckernan07} \\
& 10 & NGC 5548 & 44.30 $\pm$ 0.13 & 41.80 $\pm$ 0.26 & 7.6$^a$ & 26.35 & XSTAR & \cite{Mckernan07} \\
& 11 & MCG -6-30-15* & 43.40 $\pm$ 0.13 & 40.89 $\pm$ 0.22 & 6.5$^d$ & 26.01 & XSTAR & \cite{Mckernan07} \\
& 12 & Mrk 290 & 44.40 $\pm$ 0.13 & 42.57 $\pm$ 0.36 & 7.4$^a$ & 26.60 & Cloudy & \cite{Zhang11} \\
& 13 & Mkn 509 & 45.20 $\pm$ 0.13 & 41.93 $\pm$ 0.54 & 8.2$^c$ & 26.65 & Cloudy & \cite{Ebrero11} \\
\hline
\multirow{7}{*}{BHB} & 14 & 4U 1630 & 38.20 $\pm$ 0.43 & 32.68 $\pm$ 0.96 & 1.0$^i$ & 22.42$\pm0.30^j$ & XSTAR & this paper \\
& 15 & GRO 1655$-$40a & 37.70 $\pm$ 0.29 & 32.31 $\pm$ 0.56 & 0.83$^g$ & 21.79$\pm0.20^l$ & XSTAR & \cite{Miller08} \\
& 16 & GRO 1655$-$40b & 37.80 $\pm$ 0.29 & 33.42 $\pm$ 0.66 & 0.83$^g$ & 21.79$\pm0.20^l$ & XSTAR & \cite{Neilsen12} \\
& 17 & H 1743$-$322 a & 38.60 $\pm$ 0.46 & 33.43 $\pm$ 0.35 & 1.0$^i$ & 22.42$\pm0.30^j$ & XSTAR & this paper\\
& 18 & H 1743$-$322 b & 38.50 $\pm$ 0.43 & 32.32 $\pm$ 0.66 & 1.0$^i$ & 22.42$\pm0.30^j$ & XSTAR & this paper \\
& 19 & GRS 1915$+$105 S1 & 38.90 $\pm$ 0.13 & 33.42 $\pm$ 0.89 & 1.15$^h$ & 22.54$\pm0.03^m$ & XSTAR & \cite{Ueda09}\\
& 20 & GRS 1915$+$105 S2 & 39.50 $\pm$ 0.13 & 34.09 $\pm$ 0.33 & 1.15$^h$ & 22.54$\pm0.03^m$ & XSTAR & \cite{Miller06}\\
& 21 & GRS 1915$+$105 S3 & 39.00 $\pm$ 0.13 & 34.35 $\pm$ 0.57 & 1.15$^h$ & 22.54$\pm0.03^m$ & XSTAR & this paper \\
& 22 & GRS 1915$+$105 S4 & 39.10 $\pm$ 0.13 & 33.68 $\pm$ 0.62 & 1.15$^h$ & 22.54$\pm0.03^m$ & XSTAR & this paper \\
& 23 & GRS 1915$+$105 S5 & 39.10 $\pm$ 0.13 & 33.22 $\pm$ 0.91 & 1.15$^h$ & 22.54$\pm0.03^m$ & XSTAR & this paper \\
\hline \hline
{\bf Jets} & & & $\log L_{Bondi}$ & $\log L_{Jet}$ \\
& & & (ergs s$^{-1}$) & (ergs s$^{-1}$)\\ \hline
\multirow{12}{*}{AGN}& 24 & NGC 507 & 44.41 $\pm$ 0.09 & 44.01 $\pm$ 0.15 & 8.9$^f$ & 26.34 & - & \cite{Allen06} \\
& 25 & NGC 4374 & 43.69 $\pm$ 0.30 & 43.18 $\pm$ 0.13 & 8.8$^f$ & 25.72 & - & \cite{Allen06} \\
& 26 & NGC 4472 & 43.79 $\pm$ 0.25 & 42.91 $\pm$ 0.13 & 8.9$^f$ & 25.72 & - & \cite{Allen06} \\
& 27 & NGC 4486 & 44.16 $\pm$ 0.35 & 43.54 $\pm$ 0.23 & 9.5$^f$ & 25.72 & - & \cite{Allen06} \\
& 28 & NGC 4552 & 43.37 $\pm$ 0.22 & 42.19 $\pm$ 0.11 & 8.7$^f$ & 25.72 & - & \cite{Allen06} \\
& 29 & NGC 4636 & 42.29 $\pm$ 0.24 & 41.48 $\pm$ 0.12 & 8.2$^f$ & 25.72 & - & \cite{Allen06} \\
& 30 & NGC 4696 & 43.40 $\pm$ 0.56 & 42.90 $\pm$ 0.17 & 8.6$^f$ & 26.14 & - & \cite{Allen06} \\
& 31 & NGC 5846 & 42.85 $\pm$ 0.42 & 41.87 $\pm$ 0.16 & 8.6$^f$ & 25.88 & - & \cite{Allen06} \\
& 32 & NGC 6166 & 43.49 $\pm$ 0.30 & 43.20 $\pm$ 0.13 & 8.9$^f$ & 26.62 & - & \cite{Allen06} \\
\multirow{1}{*}{BHB}& 33 & Cygnus X-1 & 37.30 $\pm$ 0.13 & 36.00 $\pm$ 2.00 & 1.0$^i$ & 21.76$\pm0.04^k$ & - & this paper\\
\hline \hline
{\bf Ultra-Fast Outflows} \\ \hline
\multirow{3}{*}{AGN} &34& 3C 111 & 45.9 & 45.8 & 9.0 & 26.81 &XSTAR & \cite{Tombesi11} \\
&35 &APM 08279+5255 & 47.3 & 50.5 & 10.3 & 28.3 &- & \cite{Chartas02} \\
&36 &PG 1115+080 & 46.5 & 51.2 & 9.0 & 28.1 & - & \cite{Chartas07} \\ \hline
BHB & 37 & J17091+3624 & 37.5 & 38.3 & 1.0$^i$ & 22.41 & XSTAR & \cite{King12a} 
\enddata 
\label{tab:lum}
\tablecomments{* These sources have bolometric luminosities estimated from their 2--10 keV fluxes \citep[ L$_{Bol} \approx 20 L_{2-10keV}$,][]{Vasudevan09}. The masses are given by each reference unless otherwise stated; $a$: \cite{Denney10}, $b$: \cite{Lee05}, $c$: \cite{Peterson04}, $d$: \cite{Mchardy05}, $e$: \cite{Collier01}, $f$: derived from the relation given in \cite{Tremaine02} using the $\sigma$ given in \cite{Allen06}, $g$: \cite{Shahbaz99}, $h$ \cite{Greiner01}, $i$: the mass of these BHB has not been determined so a M=10$M_\odot$ and 20\% error has been assumed $j$: the distance to these sources is unknown and assumed to be 8.5$\pm4$ kpc, $k$: \cite{Reid11}, $l$: \cite{Foellmi09}, $m$: \cite{Harlaftis04}} 
\end{deluxetable*}
%-------------------------------------Table End---------------------------------------------------------
%-------------------------------------Table Start---------------------------------------------------------
\begin{deluxetable}{l l l l l}
\tablecolumns{5}
\tablewidth{0pc}
\tabletypesize{\scriptsize}
\tablecaption{Individual X-ray Wind Components}
\tablehead{ Data Set & $\alpha$ & $\beta$ & $\gamma$ & $\sigma_0$ }
\startdata
\hline Winds \\ \hline
ALL & 1.58$\pm$0.07 & -(3.19$\pm$0.19) & & 0.68 \\
BHB & 0.91$\pm$0.31 & -(5.58$\pm$1.68) & & 0 \\
AGN & 0.63$\pm$0.30 & -(1.24$\pm$0.63) & & 0.58 \\
$\log \xi > 2$ & 1.42$\pm$0.06 & -(3.73$\pm$0.14) & & 0.56\\
ALL & 0.2$\pm0.4$ & 1.2$\pm0.3$ & 24.5$\pm 0.2$ & 0.68 \\
\hline \hline Jets \\ \hline
ALL & 1.18$\pm$0.24 & -(0.96$\pm$0.43) & & 0 \\
AGN & 1.34$\pm$0.50 & -(0.80$\pm$0.82) & & 0 \\
\label{tab:para}
\tablecomments{These are the best fit parameters for each of our linear models. The $\alpha$ parameter describes the coefficient of the bolometric or Bondi luminosity, $\beta$ is the normalization of each linear fit except for the last wind fit. In that case it is the coefficient of the mass term and $\gamma$ is the normalization of the fit. Finally, $\sigma_0$ is the intrinsic scatter of each fit.}
\end{deluxetable}
%-------------------------------------Table End---------------------------------------------------------
\medskip
\medskip
\medskip
\medskip
\medskip
\medskip
\medskip
\medskip
\medskip
\medskip
\end{document}